\documentclass{aa}  
\usepackage{graphicx}
\usepackage{cancel}
\usepackage[colorlinks,allcolors=blue]{hyperref}
\bibpunct{(}{)}{;}{a}{}{,}
\usepackage{natbib}
\usepackage[varg]{txfonts}
\usepackage{tabularx} 
\usepackage{graphicx}
\usepackage{txfonts}
\defcitealias{MPK22}{MPK}

\def\apjl{\rm ApJL}
\def\apjs{\rm ApJS}

\def\aap{\rm A\&A}

\newcommand{\eqb}{\begin{equation}}
\newcommand{\eqe}{\end{equation}}

\begin{document}

   \title{Radiative feedbacks as drivers for quasi-periodic-oscillation activity in black-hole X-ray binaries}


   \author{A. Mastichiadis
          \inst{1}
          }

   \institute{Department of Physics, National and Kapodistrian University of Athens, University Campus Zografos, GR 15783, Athens, Greece\\ \email{amastich@phys.uoa.gr}}

   \date{Received March .., 2026; accepted ...}

 
  \abstract
  {Black-hole X-ray binaries (BHXRBs) in the hard and hard-intermediate spectral states commonly exhibit prominent type-C quasi-periodic oscillations (QPOs) in their X-ray power spectra. Despite extensive observational and theoretical efforts, the physical mechanism responsible for these oscillations has not yet been firmly established.}
{The disk–corona system in BHXRBs is radiatively coupled, as hard X-ray emission from the corona can be reprocessed by the accretion disk and re-emitted as soft photons that contribute to cooling the coronal electrons.   Aim of the
present study is to examine whether this feedback can give rise to limit cycles having the spectro-temporal properties of QPOs.
}
{  We model the coronal emission using a one-zone radiation framework and solve the time-dependent kinetic equations for electrons and photons. Electrons are energized by some unspecified process and cool via inverse Compton scattering of soft photons originating from (i) the accretion disk and (ii) disk reprocessing of the hard radiation produced in the corona.}
  {When electron cooling is dominated by soft photons reprocessed in the accretion disk, the disk–corona system undergoes limit-cycle oscillations. For a subset of the model parameters, these oscillations reproduce key properties of type-C QPOs observed in BHXRBs. The oscillation frequency depends  on the coronal radius and on the energization timescale, while the resulting X-ray spectra are well described by power laws extending up to energies of 
$\sim$ 10 – 100 keV. These calculations confirm and extend earlier semi-analytical results obtained with simplified treatments. Owing to the scale-invariant nature of the model, the results can be readily extrapolated to other accreting systems, such as Active Galactic Nuclei.}

                \titlerunning{QPOs in black hole X-ray binaries}
 
   \authorrunning{A. Mastichiadis}

   \maketitle
%
\section{Introduction}
BHXRBs consist of a black hole accreting matter from a low-mass companion star. 
The X-ray spectra of these objects in their hard state (for a review of the BHXRB spectral states see \citep{Belloni10}) consist of two components, a blackbody at low ($<$1\textrm{keV}) energies and a power-law that extends up to hundreds of keV energies and, in some cases, up to MeV. While the former is attributed to accretion disk emission, the latter is thought to be produced by inverse Compton scattering (ICS) of energetic electrons on the soft disk component. The electrons themselves are assumed to be energized in a region close to the central black hole, known as the corona
-- for a review see \cite{Done07}. 

Timing analysis, especially during outbursts, reveals the presence of QPOs \citep{vanderKlis89, Psaltis99, Giannios04, IngramMotta19}. Among these, the so-called type-C QPOs are characterized by a strong peak in their power spectra and appear when the source is in the hard, or hard-intermediate, state. The origin of this type of QPOs is still debated and models include, among others, instabilities in the accretion flow \citep{Tagger98}, oscillations of boundary layers \citep{Titarchuk04} or Lense-Thirring precession \citep{Stella98, Ingram09}.

\begin{figure}
\centering
\includegraphics[width = 0.45\textwidth]{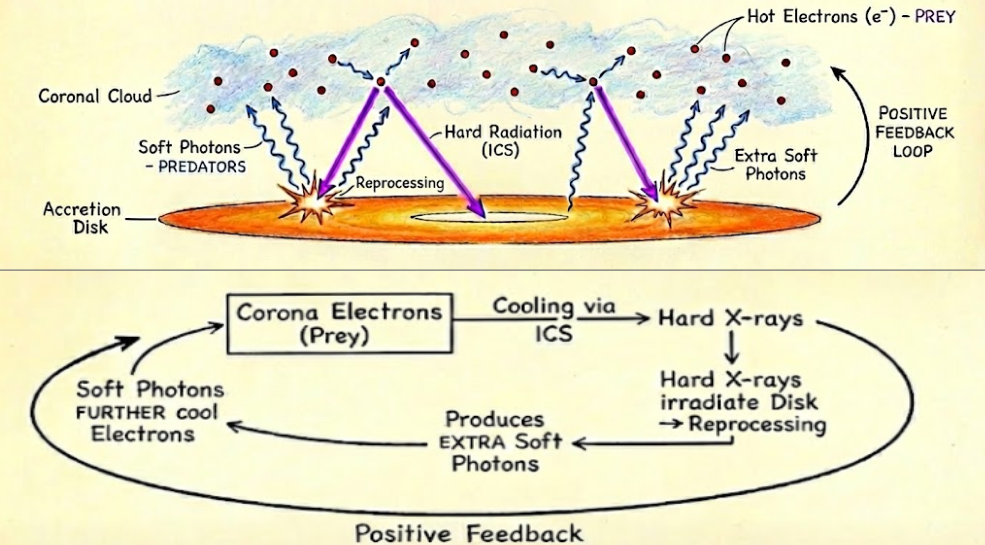}
\caption{Schematic representation  of the feedback loop between electrons and  photons in the coronal environment. For artistic reasons the corona is depicted as a cloud, while in the paper its shape is assumed to be spherical. (The figure is AI generated.) }
\label{model}
\end{figure}

In a recent work, \cite{MPK22} (henceforth \citetalias{MPK22}) showed that interactions between the corona electrons and the accretion disk soft photons can be inherently non-linear, giving rise to an oscillatory pattern in the X-ray flux, reminiscent of limit cycles found in non-linear dynamical systems. Their model was a dynamical treatment of \cite{Haardt91}: The soft photons of the accretion disk cool by ICS  the corona electrons $\rightarrow$ Part of the electron ICS hard radiation is reprocessed on the accretion disk, producing extra soft photons there $\rightarrow$ The extra soft photons further cool the electrons. 
This succession of steps forms a positive feedback loop between the electrons and soft photons as illustrated in Fig.~\ref{model}. If the energetic electrons are continuously replenished,
the modeling of the electron-photon coupling leads to a type of Lotka-Volterra, or prey-predator, system of equations with electrons being the prey and the soft photons the predators. The solution of this type of equations leads to limit cycles. 
\citetalias{MPK22}'s treatment was simplified in the sense that they used the energy integrated densities for the two species. Therefore, their results were limited to the temporal properties of the system; yet, their approach was self-consistent and kept the physical essence of the interactions. 

In the present paper, we extend the above approach by considering the 
full spectro-temporal evolution of the system. This essentially means that the set of the two coupled differential equations that described the system in the case of \citetalias{MPK22} has to be replaced by a large set of coupled partially differential equations that describe the simultaneous evolution in time and energy of the electron and photon distribution functions. In order to solve these equations we have used a numerical code that self-consistently treats
 the relevant processes between electrons and photons 
\citep{MastichiadisKirk95, Dimitrakoudis12}.
Electrons are assumed to be energized 
inside the corona
by some (unspecified) mechanism and, at the same time, lose energy on soft photons. These are not independent of the electrons but come from reprocessing of a part of the electron ICS radiation on the accretion disk. This coupling of electron energization and cooling can give QPOs, as a result of the limit cycle behavior, to the system, exactly as in \citetalias{MPK22}. However, in this case one can get more information because the numerical code gives the electron and photon spectrum at every single time instant. 


The aims of the present paper are (i) to show that limit cycles are inherent in a corona-disk system and (ii) to make a broad study of the physical parameters that can give results resembling the spectral and temporal behavior of the BHXRBs in their hard X-ray state. 
It is structured as follows. In Sect.~\ref{sec:2}  we outline the model and present the equations that describe the interaction between the electrons in the corona and the soft radiation from the disk. In Sect.~\ref{sec:3}  we present briefly the specifics of the code used. In Sect.~\ref{sec:4} we give a typical example of the numerical solutions to the equations of the corona-disk system and define some temporal and spectral parameters that characterize these solutions. In Sect.~\ref{sec:5} we systematize our results by investigating the effect of each of the main parameters of the problem.  In Sect.~\ref{sec:6} we construct and solve analytically  a simplified set of equations showing explicitly why the non-linearity arises.
In Sect.~\ref{sec:7} we give an application on the system GRS 1915+105 and we conclude in Sect.~\ref{sec:8} with a summary and a discussion.

\section{Model description}\label{sec:2}

We start by adopting the standard picture of an accretion disk - corona system. The exact geometry is not of relevance here; all we require is that the disk provides soft photons for the cooling of the electrons of the corona. Electrons enter the corona, which is a spherical region of radius $R_{c}$, where they are energized by some unspecified mechanism. For the purpose of the present work it is sufficient to attribute to this mechanism only a characteristic energization timescale $ t_{en}$ that is assumed to be independent of the electron energy for simplicity. Here we will adopt the approach of \cite{KRM98} -- see also \cite{Petropoulou24} for a recent application to non-thermal flares from  SgrA$^*$.

In the absence of losses, the electron Lorentz factor obeys the characteristic equation,
\begin{equation}
\frac{d\gamma}{dt}=\frac{\gamma} {t_{en}},
\end{equation}\label{eq:1}
which, for initial condition $\gamma(t_0)=\gamma_{0}$, gives 
\begin{equation}
\gamma(t)=\gamma_{0}e^{(t-t_0)/ t_{en}}.
\end{equation}\label{eq:2}
As the electrons are energized, they start losing energy through various processes. If we concentrate on ICS, then we can write the characteristic equation for the electron energy as
\begin{equation}\label{eq:3}
\frac{d\gamma}{dt}={\frac{\gamma}{ t_{en}}}-\frac{4}{3}\sigma_Tc\left(\frac{U_{\rm s}}{m_ec^2}\right)\gamma^2,
\end{equation}
where $\sigma_T$ the Thomson cross-section, $m_e$ the electron rest mass and $U_s$ the soft photon energy density. Note that while for $U_s$=const, the above equation leads to an equilibrium at 
\begin{equation}
\gamma_{sat}=\frac{3m_ec}{4\sigma_TU_s t_{en}},
\end{equation}
 a steady state might not be established in the case $U_{\rm s}=U_{\rm s} (t).$ 

To describe, therefore, the behavior of electrons in time and energy inside the corona we assume 
that these enter homogeneously the region with initial Lorentz factors $\gamma_{inj}$  at a rate $q_{inj}$ per unit volume. Once inside the corona the electrons can be energized with the timescale $ t_{en}$, lose their energy by ICS
or physically escape the source
on a timescale $t_{esc}$. Thus, in the case where ICS is the sole energy loss mechanism, the temporal evolution of the differential density electron distribution, $n_e(\gamma, t)$ (units $\textrm{volume}^{-1}\textrm{energy}^{-1})$, is described by a partial differential equation \citep{KRM98}:
\begin{equation}
\frac{\partial n_e(\gamma,t)}{\partial t} + \frac{\partial}{\partial \gamma}\left[ \left(\frac{\gamma} {t_{en}} - b_c \gamma^2\right) n_e(\gamma,t) \right] + \frac{n_e(\gamma,t)}{t_{esc}} = q_{inj}\delta(\gamma-\gamma_{inj}),
\label{eq:kinetic5}
\end{equation}
where $b_c = 4\sigma_T  U_s/ (3 m_e c)$ and $\delta(x)$ is the Dirac delta function. In the case where $ t_{en}$ and $t_{esc}$ are independent of the electron energy, $q_{inj}$ does not vary in time and $\gamma$ is far from the upper cutoff of the distribution, the solution is described by a power-law, i.e.
$n_e(\gamma)\propto \gamma^{-1-(t_{en}/t_{esc})}$ which gets asymptotically to $-1$ as $t_{esc}\rightarrow \infty$. In such cases, most of the total electron energy content is carried by the high energy end of the electron distribution and this will be one of our assumptions in the present paper. 

Eq.~\ref{eq:2} implies that the electron cooling term produces radiation and
should be coupled by an analogous equation for photons that describes their spectral and temporal properties inside the corona. Assuming again that ICS is the main mechanism for electron radiation, the equation for photons can be written as follows. 
\begin{equation}\label{eqkin:6}
\frac{\partial n_\gamma(\epsilon,t)}{\partial t} + \frac{n_\gamma(\epsilon,t)}{t_{cr}} = \mathcal{Q}_{ICS}(\epsilon,t), 
\end{equation}
where $n_\gamma(\epsilon,t)$ is the differential photon density (units volume$^{-1}$energy$^{-1})$, $\epsilon$ is the photon energy, and $t_{cr}=R_c/c$ is the crossing time of the corona. The term $n_\gamma/t_{cr}$ denotes the photon escape while $\mathcal{Q}_{ICS}$ is the IC emissivity -- see, e.g., \cite{BG70}.

Eqs \ref{eq:kinetic5} and \ref{eqkin:6} are coupled because $\mathcal{Q}_{ICS}$, which appears in the photon equation, is a function of $n_e(\gamma,t)$. Furthermore, the solution of the electron equation depends on the photon energy density $U_s$, which for the corona-disk case might depend on $n_e$, thus making the problem non-linear. Therefore, in order to calculate $U_s$, one has to make some further assumptions about its origin. Here we follow the approach of \citetalias{MPK22} and assume that there are two components contributing to $U_s$:
(i) the disk thermal radiation of temperature $T_{eff}$, which is assumed to be constant and (ii) the reprocessed hard radiation from the corona impinging on the disk, which depends on the ICS component calculated from Eq.~\ref{eqkin:6}. Therefore we can write
\begin{equation}\label{equation7}
    U_s=U_{ds}+U_{rs},
\end{equation}
where $U_{ds}$ is the energy density of the thermal radiation of temperatude $T_{d}$ coming from the disk and $U_{rs}$ is the energy density of the reprocessed radiation.
For the latter, we assume that it also emits as a gray body of temperature $T_{eff}$ and, in order to avoid the use of many parameters we set $T_{eff}=T_{d}$. The energy density of the reprocessed component is calculated at each instant by the relation
\begin{equation}\label{equation8}
U_{rs}(t)=\alpha\int d\epsilon\epsilon n_\gamma(\epsilon,t),
\end{equation}\label{eq:8}
with the integration taking into account only the hard photons of energies $>k_BT_{eff}$ while $\alpha$ is the fraction of the hard corona radiation that is reprocessed by the disk.  

The above describe both the parameters and the assumptions of the problem and, therefore, one can solve the electron and photon equations in a straightforward manner to obtain their tempo-spectral behavior. In the next section, we 
give some details about the numerical code used and also relate the parameters to the physics of BHXRBs.

\section{Code specifics}\label{sec:3}

The numerical code used is based on the leptonic part of the time-dependent code $ATHE\nu A$ \citep{MastichiadisKirk95, Dimitrakoudis12} which takes into account mostly nonthermal processes. For an updated description and comparisons with other, similar, codes, see \cite{Cerruti24}.
The code has been modified in two ways: (a) it uses an explicit electron energization scheme as used in 
 \cite{KRM98} and \cite{Petropoulou24}, rather than an electron injection term in high energies as is the norm with this type of codes, and (b) as explained in the previous section, it uses reprocessed photons for ICS cooling. This means that at each time step the ICS emissivity is integrated, turned into radiation density  and a fraction $\alpha$ of it is assumed to be radiated as a gray body of temperature $T_{eff}$. 
 The energy snapshots have been recorded in time steps $\Delta t=0.1 t_{cr}$. The physical processes used are (i) inverse Compton scattering, (ii) photon-photon pair production, and (iii) electron-proton bremsstrahlung. It turns out that for the majority of the parameters used, the impact of the two latter processes is minimal; however, they have been used throughout for completeness reasons. In contrast, synchrotron radiation that can potentially have an effect on our results is not taken into account; the implications that this has on the value of the B-field will be discussed in Sect.~\ref{sec:6}

 
The free parameters of the problem are the following.
\begin{itemize}
\item 
The corona radius $R_c$.
\item 
The electron energization timescale $ t_{en}$.
\item
The electron injection rate $q_{inj}$ or, equivalently, the mass accretion rate $\dot M_{inj}$ (see below).
\item 
The electron escape timescale $t_{esc}$.
\item 
The accretion disk thermal luminosity. 
\item 
The temperature of the disk and the  reprocessed component $T_{eff}$, assumed to be equal.
\item 
The fraction $\alpha$ of the hard radiation that is reprocessed by the disk and is radiated back as soft photons.  
\end{itemize}

We note
that most of the above parameters have estimates in the BHXRB literature. Only $ t_{en}$ and $t_{esc}$ are practically unknown, therefore we will use generic values for them that are simple multiples of the crossing timescale $t_{cr}=R_c/c$ and we will use $ \tilde t_{en}= t_{en}/t_{cr}$.
For simplicity we will set $\tilde t_{esc}=10^3\tilde t_{en}$, i.e. we will require that the electron distribution function is flat, carrying most of the electron energy close to its upper cutoff.  Other values of the ratio $t_{esc}/t_{en}$ will be discussed at the end of Sect.~\ref{sec:5}.

For the electron injection term at low energies (see Eq.~\ref{eq:kinetic5}) we will use $\gamma_{inj}=10^{0.05}$. Electrons with $\gamma<\gamma_{inj}$ will be considered as cold. Note that 
the injection rate of electrons per unit volume $q_{inj}$, used in the same equation,
does not have a straightforward connection to some observable quantity as it gives the rate of low energy electrons entering the energization process. However, if one assumes an electron-proton plasma, they can write for the total injection rate 
\begin{equation}\label{eq:qinj}
Q_{e,inj}\simeq \frac{4\pi}{3}R_c^3q_{inj}m_ec^2=\xi\frac{\dot {M}_{inj}}{m_p},
\end{equation}
where $\dot{M}_{inj}$ is the mass accretion rate, $m_p$ is the proton mass and $\xi$ is the fraction of electrons that enter into the energization process and which, for simplicity, we will assume to be equal to 1. Note that Eq.~\ref{eq:qinj} does not imply a relation of $\dot M_{inj}$ to the luminosity of the system. We will come come back to this point in the next section.

Finally, since the system of Eqs~\ref{eq:kinetic5} and \ref{eqkin:6} is time-dependent, we will use as initial conditions $n_e(\gamma,0)=n_\gamma(\epsilon,0)=0$ everywhere with the exception of the electrons first bin at 
$\gamma_{inj}=10^{0.05}$ which takes the boundary value $n_e(\gamma_{inj},0)=q_{inj}t_{en}$ -- see  \cite{KRM98}.

In the next section, we give a typical example of how the system moves from linear to nonlinear behavior and use this to define some useful quantities that we will use extensively for the rest of the paper. However, a detailed description of the mechanisms behind the limit cycle behavior can be found in Appendix~\ref{app1}.

\section{The effect of disk photons on the temporal evolution of the system}\label{sec:4}

We show below some results that represent the typical features of our model. In the next  section we will use normalized quantities.  So the hard luminosity, the disk soft energy density and the reprocessed on the accretion disk energy density will be given in terms of compactnesses,
\begin{equation}
\ell_h=\frac{L_h\sigma_t}{4\pi R_c m_ec^3},
\end{equation}
\begin{equation}
\ell_{ds}=\frac{U_{ds}\sigma_TR_c}{m_ec^2},
\end{equation}\label{eq:10}
\begin{equation}
\ell_{rs}=\frac{U_{rs}\sigma_TR_c}{m_ec^2},
\end{equation}
respectively, where
$L_h$ the hard luminosity produced from ICS.
Furthermore, time will be measured in light crossing time of the corona. The photon energy will be given in keV.

\begin{figure}
\centering
\includegraphics[width = 0.35\textwidth]{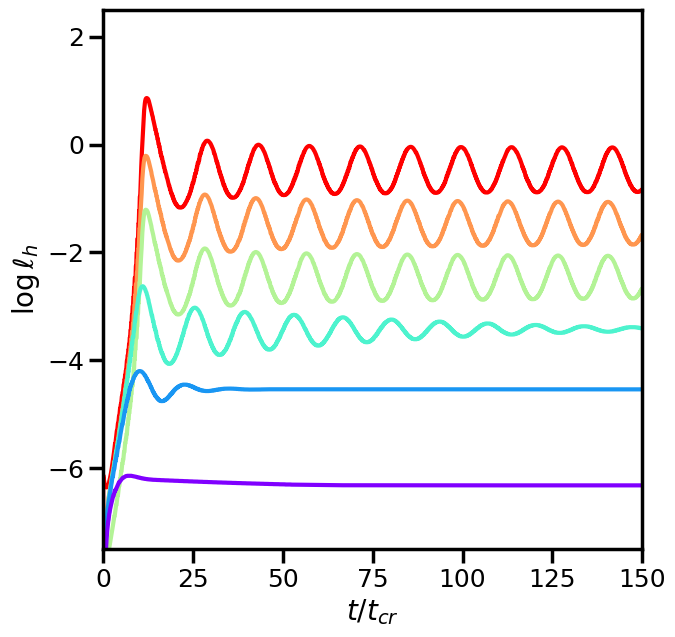}
\caption{Light curves of ICS compactness in the 2-200 keV band for various values of the parameter $\ell_{ds}$. All other parameters have been kept constant and have the values
$R_c=10^9~\textrm{cm},~\dot M_{inj}=3.2\times10^{-7}M_\odot/\textrm{yr},~\tilde t_{en}=2.5,~t_{esc}/t_{en}=10^3,~T_{eff}=4\times 10^6~\textrm{K},~\alpha=0.1$. The colors of the lines correspond to $\ell_{ds}=10^{-6},~10^{-5},~10^{-4},~10^{-3},~10^{-2}$ and $10^{-1}$ from red to violet. For the sake of clarity each curve is displaced from the previous one by unity.}
\label{f_light}
\end{figure}

As a first result, we show the effects of soft disk radiation, expressed through $\ell_{ds}$, on the temporal behavior of $\ell_h$.  Figure~\ref{f_light} shows a plot of $\ell_h$ versus normalized time for various values of $\ell_{ds}$ with all other parameters  fixed.
As $\ell_{ds}$ is decreasing, the system becomes nonlinear and limit cycles start appearing -- see the succession of light curves from violet to red. Generally speaking, as long as $\ell_{ds}<<\ell_{rs}$, the cooling occurs in the non-linear regime.
From the obtained light curves 
various quantities can be extracted, like:
\begin{itemize}
\item 
The frequency of the QPO $\nu_{QPO}$. This is calculated directly from the light curves by using standard python routines. Here and for the rest of the paper, for computational resources reasons, we have used the first ten cycles of the light curves, excluding the first one in order to minimize the effects of the initial conditions.
Despite the fact that $\nu_{QPO}$ can be calculated for the whole hard spectrum, here 
we have chosen the energy range 2-200 keV in order to keep an analogy to the observations that are usually performed in this energy range. 
\item 
The fractional rms of the QPO $f_{rms}$. 
This was calculated for the same section of the light curves  and for the same energy range as above.
\item
The timescale $\tilde T_{10}$, expressed in units of $t_{cr}$, required for the X-ray light curve to drop to $10\%$ of its value at the second peak.
As the damping of the obtained light curves does not fit any simple analytical law, we resorted to the calculation of $\tilde T_{10}$. 
For this we have used the expression for the difference norm
\begin{equation}
D(\tilde t)=\left|n_\gamma(\epsilon_i,\tilde t+\Delta\tilde t)-n_\gamma(\epsilon_i,\tilde t)\right|,
\end{equation}
and have numerically determined $\tilde T_{10}$ from the equation
\begin{equation}
D(\tilde T_{10})=0.1 D(\tilde t^{pk}_2),
\end{equation}
where $D(t^{pk}_2)$ is the value of the norm at the second peak of the light curve. Here $\Delta \tilde t=0.1$ and by $\tilde t$ we mean that time is normalized to $t_{cr}$.
\item 
The phase lag $\Delta\phi$ between two characteristic energy bands. Here we have chosen the 2-6 keV as the "soft" energy band and the 6-15 keV as the "hard" energy band in line with most of the QPO literature -- e.g. \cite{Zhangetal20}. We have calculated the time lag $\Delta T_{sh}$ between the two bands 
and, consequently, we have calculated the corresponding phase-lag by using the relation $\Delta\phi=2\pi\nu_{QPO}\Delta T_{sh}$.
 In the present work a positive lag means that the
hard photons lag the soft ones. A detailed description on the behavior of $\Delta T_{sh}$ can be found in Appendix~\ref{app2}
\end{itemize}

The so-calculated quantities are shown in Tab.~\ref{tab:1} as a function of the compactness of the disk $\ell_{ds}$  -- here, in addition to $\Delta\phi$, we have tabulated $\Delta T_{sh}$ (in units of $t_{cr}$) as well, as this is the quantity that is directly computed from the numerical code. Several conclusions can be drawn that, despite covering only one set of parameters, have more general consequences. 

\begin{itemize}
\item 
As already mentioned above, as
$\ell_{ds}$ is decreasing, the system becomes nonlinear and limit cycles start appearing. Generally speaking, as long as $\ell_{ds}<<\ell_{rs}$, the cooling occurs in the non-linear regime. We also note that once established, the period of the oscillations remains largely unaffected by the value of $\ell_{ds}$, i.e. there is a limit that the system tends to reach which is its {\sl{natural}} frequency.
The lower the value of $\ell_{ds}$, the stronger the limit cycles. This result verifies the findings of \citetalias{MPK22} that QPOs appear when the disk is not radiating but acts, instead, as a reproccessor of the hard ICS radiation. On the other hand, a luminous disk damps very effectively the limit cycles as the plentiful, in this case, disk photons provide efficient cooling for the corona electrons linearizing, at the same time, the whole process.
\item 
In contrast to \citetalias{MPK22} that found only damped oscillations, here we find cases that are, practically speaking, undamped. We will explain this difference in a next section.
\item
The quantity $\tilde T_{10}$ is a good alias for the damping of the system. Generally speaking, runs that show $\tilde T_{10}<50$ are strongly damped. On the other hand, runs with $\tilde T_{10}>500$ can practically be considered as undamped.
\item
In contrast to $\nu_{QPO}$, which is rather insensitive to damping, fractional rms is a sensitive function of it. Therefore, the values shown in the table for $\tilde T_{10}<100$, where damping starts becoming stronger, can only be considered as upper limits. 
\item 
The quantity $\Delta T_{sh}$ that measures soft/hard lags is rather independent of $\ell_{ds}$. Here it is negative, meaning that the
soft photons lag the hard ones.
\end{itemize}

\begin{table}[h]
  \centering
  \caption{Characteristic values of the QPO frequency $\nu_{QPO}$, the fractional rms $f_{rms}$, the 10-fold decay timescale $\tilde T_{10}$ and the time difference between soft and hard energy bands $\Delta T_{sh}$ for various values of $\ell_{ds}$.}
  \begin{tabular}{c c c c c c}
\hline \hline
   $\ell_{ds}$ & $\nu_{QPO}$~(Hz) & $f_{rms}$~(\%) & $\tilde T_{10}$ & $\Delta$ $T_{sh}$/$t_{cr}$ & $\Delta\phi$ \\
   \hline
   $10^{-6}$  & 2.08 & 20.4 & 2500 &  -0.30 &-0.13\\ 
    $10^{-5}$ & 2.09 & 20.3 & 2000 &  -0.30 &-0.13\\
    $10^{-4}$  & 2.11 & 20.2 & 1400 & -0.30 &-0.13\\ 
    $10^{-3}$ & 2.29 & 10.8 & 1200 & -0.30 &-0.14\\ 
    $10^{-2}$  & 2.52 & 0.9 & 20 & -0.30 &-0.16\\ 
    $10^{-1}$  & N/A & N/A & N/A & N/A &N/A\\ 
    \hline

  \end{tabular}
    \tablefoot{    The above quantities have been averaged between the second and tenth peaks of the photon light curve.
    In the last column the minus sign means that for the particular example soft energies lag hard. For the input parameters used see text.}
  \label{tab:1}
\end{table}

\begin{figure}
\centering
\includegraphics[width = 0.53\textwidth]{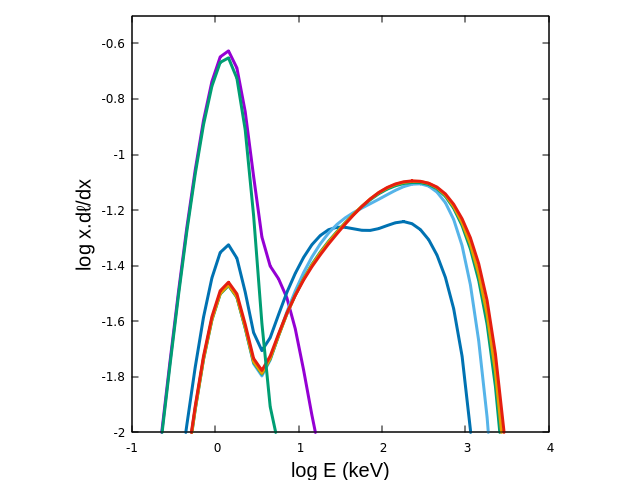}
\caption{Average photon spectra for the same cases shown in the previous figure. The color codes have been kept the same.}
\label{f_MW}
\end{figure}

Figure~\ref{f_MW} shows the averaged, over the first ten cycles,  photon spectra for the same cases, as above. In these plots, one can see the superposition of the disk and reprocessed soft photons, which have been assumed to be gray bodies of temperature $T_{eff}=4\times10^6~\textrm{K}$, and the inverse Compton component that reaches, in this particular example, MeV energies. Here, the $y-$axis is in  units $x\frac{d\ell}{dx}=x^2\dot{n}_\gamma(x,t)$, 
where $x=\epsilon/m_ec^2$.
This is equivalent to $\epsilon\frac{dL_\gamma}{d\epsilon}$
where $L_\gamma$ is the spectral luminosity.

As one can observe, low values of $\ell_{ds}$  produce 
harder ICS spectra; however, as $\ell_{ds}$ increases, the disk photons hinder electron acceleration, and as a result both electrons  is lower and  ICS photons reach lower energies (see blue line curve). Ultimately, in the case of strong disk emission, electrons
cannot, practically speaking,
 be energized and, therefore, the photon spectrum consists only of the input disk gray-body photon distribution (violet line curve). 

From the spectra shown in Fig.~\ref{f_MW} various quantities can be extracted. 
\begin{itemize}
\item 
The bolometric inverse Compton luminosity, $L_{bol}$.
\item 
The X-ray luminosity between 2 and 200 keV, $L_X$.
\item
The peak of the ICS component, $E_{pk}$ -- Fig.~\ref{f_MW} is in units such that it immediately shows this quantity. 
\item 
The slope of the photon spectrum between 2-200 keV.
\item
The efficiency of the acceleration process  
$\eta= \dot{M}_{eff}/\dot{M}_{inj}$,
where $\dot{M}_{eff}=L_{bol}/c^2$.
This quantity signifies the fraction of injected rest mass that ultimately goes into radiation.  
\item 
The optical depth of the cooled electrons $\tau_T$. This is the optical depth of the electrons that have cooled by ICS at Lorentz factors $\gamma<\gamma_{inj}$ including the unenergized ones of the flow.
\end{itemize}

Table~\ref{tab:2} shows the above quantities, with the exception of $\tau_T$ which remains in most cases below $0.1$ signifying an optically thin medium\footnote{The Thomson optical depth when all electrons (cold and energetic) are used does not exceed 0.3.}. All remain 
practically constant for low values of $\ell_{ds}$; however, as $\ell_{ds}$ and the degree of damping increases, the luminosity drops and the spectrum steepens.

\begin{table}[h]
  \centering
  \caption{Characteristic values of the average bolometric luminosity $L_{bol}$, the X-ray luminosity (between 2-200 keV) $L_X$, the energy of the peak of the spectral luminosity $E_{pk}$, the photon spectral index $\Gamma$ between 10-100 keV and the efficiency $\eta$ as defined in Sect.~\ref{sec:4}. }
  \begin{tabular}{ c c c c c c c}
    \hline\hline
  $\ell_{ds}$ & $L_{bol}$~(erg/s) & $L_X$~(erg/s)  & $E_{pk}$ $~(keV)$ & $\Gamma$ &$\eta$  \\ \hline
    $10^{-6}$ &$3.2\times 10^{38}$& $1.0\times 10^{38}$ & 230 &  1.6 & 0.0085 \\ 
    $10^{-5}$ & $3.2\times 10^{38}$& $1.0\times 10^{38}$ & 230 &  1.6 & 0.0085  \\ 
   $10^{-4}$ & $3.2\times 10^{38}$& $1.0\times 10^{38}$ & 230 &  1.6 & 0.0085 \\ 
   $ 10^{-3}$ & $2.9\times 10^{38}$& $1.0\times 10^{38}$ & 290 &  1.6 & 0.0076 \\ 
   $ 10^{-2} $& $2.7\times 10^{38}$&$ 1.0\times 10^{38}$& 180 &  1.8 & 0.0071 \\ 
   $ 10^{-1}$ & $2.0\times 10^{38}$& $0.7\times 10^{38}$ & N/A &  N/A & 0.0071 \\ \hline   
  \end{tabular}
  \tablefoot{  The above quantities have been averaged between the second and tenth peaks of the light curve.
  For the input parameters used, see the caption of Fig.~\ref{f_light}}
  \label{tab:2}
\end{table}

\section{General trends}\label{sec:5}

According to the findings of the previous section, by assuming a cold disk, we are maximizing the chance of limit cycle appearance.
Therefore, by fixing the disk photon energy density at some low value, i.e. $\ell_{ds}=10^{-6}$, we will search for QPO appearance by changing three key parameters. For this we will first obtain light curves for various values of $\tilde t_{en}$ while keeping all other parameters fixed. Once we detect limit cycle behavior we will calculate the quantities defined in the previous section and given in 
Tables~\ref{tab:1} and \ref{tab:2}
As a next step, we will repeat the same procedure, this time by changing $\dot M_{inj}$ and, finally, we will consider variations in  $R_c$. 
At the end of the section we will discuss briefly variations in auxiliary parameters like the reprocessing factor $\alpha$, the reprocessing temperature $T_{eff}$ and the ratio $t_{esc}/t_{en}$.

\subsection{Energization timescale ${\tilde t_{en}}$ variations}\label{sec:5.1}

As we have not adopted any specific theory behind the electron energization process in the corona, 
we follow a generic approach and vary the energization timescale $\tilde t_{en}$ from fast ($\tilde t_{en}=1.25$) to slow 
($\tilde t_{en}=80$) using a multiplication step of 2 between successive runs. We note that the in the context of the one-zone model we require $\tilde t_{en}>1$.

\begin{figure}
\centering 
\includegraphics[width =0.4\textwidth]{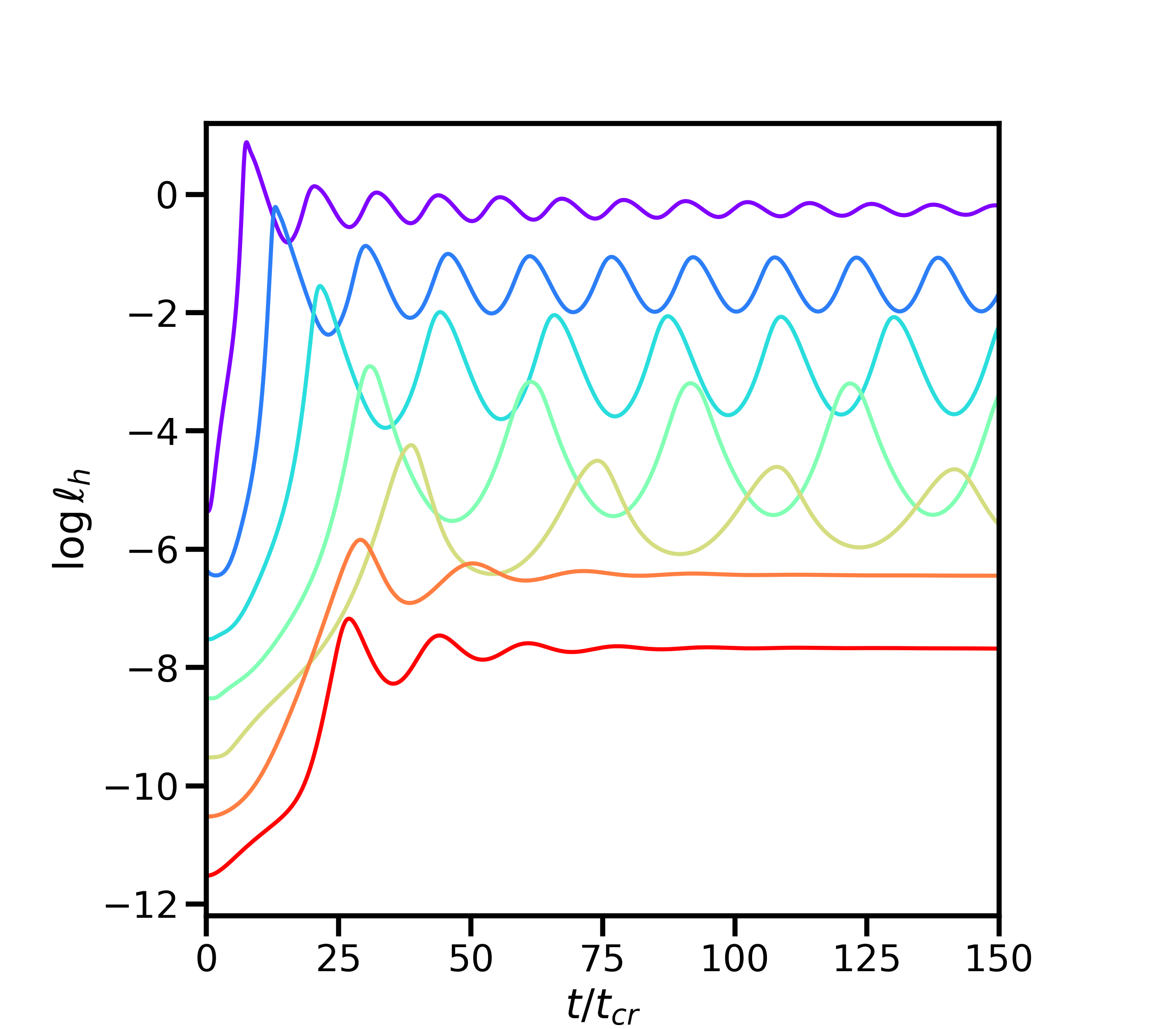}
\caption{Plot of a set of light curves all sharing the same value of 
$Q_{e,inj}$ ($M_{inj}=16\times10^{-8}M_\odot/\textrm{yr} $) but having different energization timescales. Here $\tilde t_{en}=1.25$ (violet), $2.5$ (dark blue), $5$ (light blue), $10$ (green), $20$ (yellow), $40$ (orange) and $80$ (red), all given in units of $t_{cr}$. The other parameters are 
$R_c=10^9~\textrm{cm},~t_{esc}/t_{en}=10^3,~\alpha=0.1,~T_{eff}=4\times 10^6~\textrm{K},~ \ell_{ds}=10^{-6}$. For the sake of clarity each curve is displaced from the previous one by 1. }
\label{figten}
\end{figure}

\begin{figure}
\centering
\includegraphics[width = 0.35\textwidth]{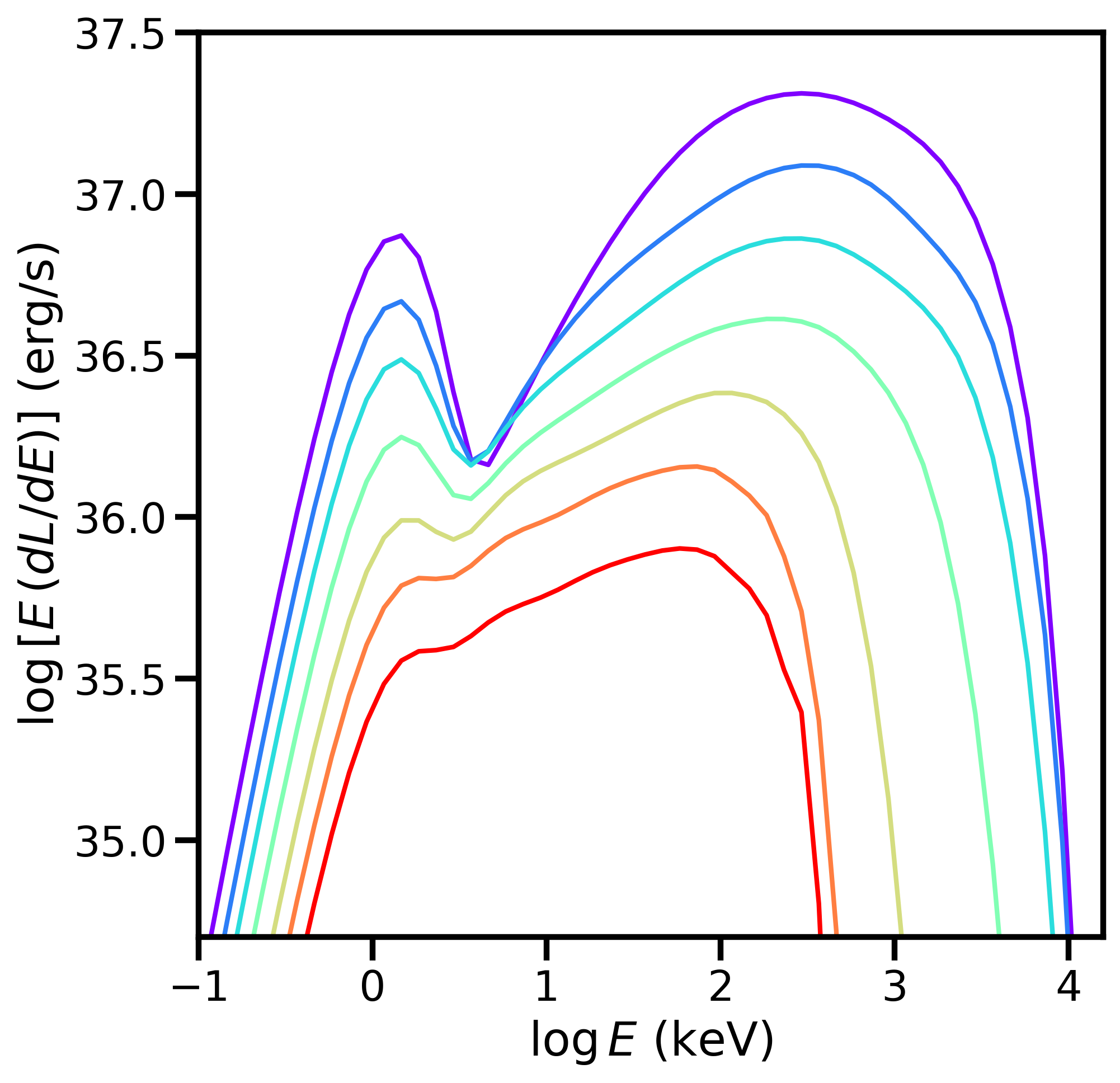}
\caption{Plot of the average multiwavelength spectrum for the case shown in the previous figure keeping the color codes the same.
}
\label{figtenMW}
\end{figure}

Figures~\ref{figten} and \ref{figtenMW} show, respectively, the light curves and averaged photon spectra produced for the above described values of $\tilde t_{en}$ when $\dot M_{inj}=16\times 10^{-8} M_\odot/\textrm{yr}$ and $R_c=10^9~\textrm{cm}$, while Tab.~\ref{tab:3} tabulates the parameters defined in the previous section. The parameters for these runs are given in the captions of Fig.~\ref{figten} and Tab.~\ref{tab:3}.

\begin{table*}[h]
  \centering
  \caption{Results as a function of $\tilde t_{en}$.}
  \begin{tabular}{c c c c c c c c c c c}
    \hline\hline
    $\tilde t_{en}$ &
    $\nu_{QPO}(Hz)$ & $\tilde T_{10}$ & $f_{rms} (\%)$ & $\Delta T_{sh}/t_{cr}$ & $\Delta\phi$ & $L_{bol}(erg/s)$ & $L_X(erg/s)$ & $E_{pk}(keV)$ & $\Gamma$ & $\eta$  \\ \hline
    1.25 & 2.59 & 180 & 13.7 & 0.53 & 0.29 & $2.8\times 10^{38} $& $1.3\times 10^{38}$ & 360 & 1.1 & 0.018 \\ 
    
    2.5 & 1.95 &1500 & 23.5 & -0.04 & -0.02 &$ 1.7\times 10^{38}$ & $8.5\times 10^{37}$ & 280 & 1.2 & 0.011 \\ 
    
    5 & 1.39 & 5000 & 31.8 & -0.30 & -0.09 &$ 1.1\times 10^{38} $& $5.9\times 10^{37} $& 230 & 1.5 & 0.0067 \\ 
    
    10 & 0.99 & 10000 & 29.3& -0.39 & -0.08 & $5.9\times 10^{37}$ & $3.8\times 10^{37}$& 180 & 1.8 & 0.0035 \\ 
    
    20 & 0.89 & 850 & 21.0 & -0.48 & -0.09 & $2.9\times 10^{37} $& $2.5\times 10^{37}$& 120 & 1.9 & 0.0018 \\ 
    
    40 & 1.49 & 50 & 6.7 & -0.44 & -0.14 & $1.5\times 10^{37}$ & $1.5\times 10^{37}$& 70 & 1.9 & 0.00098 \\ 
    
    80 & 1.70 & 40 & 5.8 & -0.33 & -0.12 & $8.8\times 10^{36} $&$ 8.8\times 10^{36}$ & 60 & 1.9 & 0.00057 \\ \hline

  \end{tabular}
  \tablefoot{Table showing the values of the various calculated parameters of the problem -- for definitions see captions of Tables 1 and 2) as a function of $\tilde t_{en}$.
   Other initial parameters are $R_c=10^9~\textrm{cm},~\dot M_{inj}=16\times10^{-8}M_\odot/\textrm{yr},~t_{esc}/t_{en}=10^3,~\alpha=0.1$ and $T_{eff}=4\times 10^6~\textrm{K}$.
  
  }
  \label{tab:3}
\end{table*}

There are several conclusions that can be deduced from these runs.
\begin{itemize}
\item 
QPOs appear for all values of $\tilde t_{en}$ with strong damping becoming evident only for large values of the energization timescale, in the present example for $\tilde t_{en}>40$. Therefore, according to the discussion in Sect.~\ref{sec:4}, these cases are inherently damped, i.e. their damping does not depend on the level of  disk radiation.
However, as we will show in the next subsection, strong damping can occur at different values of $\tilde t_{en}$ for other values of $\dot{M}_{inj}$.
\item 
The QPO frequency $\nu_{QPO}$ decreases with increasing $\tilde t_{en}$ up to $\tilde t_{en}=20$, then it starts increasing again. However, as mentioned above, the runs with $\tilde t_{en}>40$ produce heavily damped oscillations, so these cases can be considered as outliers.
\item
 Faster acceleration timescales tend to produce spectra, which (i) have higher bolometric luminosities,  (ii) peak at higher photon energies and (iii) have harder spectral indices $\Gamma$. Note that 
 $L_{bol}$, and therefore $\eta$, increases roughly linearly with $\tilde t_{en}^{-1}$. On the other hand, 
 $L_X$ changes much less than $L_{bol}$ because as the photon spectra are hard, most of the luminosity is carried around the peak of the photon distribution, $E_{pk}$, that lies above the band 2-200 keV used to calculate $L_x$. However, for slow $\tilde t_{en}$, $E_{pk}$ is decreasing and $L_{X}$ becomes eventually  equal to $L_{bol}$. 
\item 
The fractional rms $f_{rms}$ is of the order of tens of percent. However, these values must be considered strictly as upper limits, especially in those cases that are characterized by strong to intermediate damping. Generally, $f_{rms}$ and $\tilde T_{10}$ have an erratic behavior that will be discussed further down.
 \item 
 The quantity $\Delta T_{sh}$ used to measure soft/hard lags decreases with increasing $\tilde t_{en}$ and ultimately changes sign, therefore the present model can exhibit both soft and hard lags depending on the initial parameters. An explanation of this is given in Appendix~\ref{app2}.  
 \item 
 Both Fig.~\ref{figtenMW} and Tab.~\ref{tab:3} show that the X-ray spectra are hard ($\Gamma<2$). This does not hold for all cases, however. For instance, cases with higher values of the reprocessing parameter $\alpha$ tend to produce steeper spectra. 

\end{itemize}

\subsection {Mass accretion rate $\dot {M}_{inj}$ variations}

We can repeat the procedure described above for different values of $\dot{M}_{inj}$, 
while keeping the other parameters constant. Indicative results are shown in Tab.~\ref{tab:4}. We note that, contrary to the previous case, $\nu_{QPO}$ remains largely independent of the value of $\dot{M}_{inj}$, while both $L_{bol}$ and $L_X$ tend to increase with $\dot{M}_{inj}$. However, in the case of $L_{bol}$ the increase is very slow, i.e., despite the fact that $\dot{M}_{inj}$ changes by more than two orders of magnitude (a factor of $2^7=128$ to be exact), $L_{bol}$ changes only by about a factor of 3. This has the result that the efficiency $\eta$ is a decreasing function of $\dot{M}_{inj}$. In the case of $L_X$ the change is greater, by a factor of about 30, because $E_{pk}$ increases as $\dot{M}_{inj}$  decreases.

\begin{table*}[h]
  \centering
  \caption{Results as a function of the mass accretion rate $\dot{M}_{inj}$.}
  \begin{tabular}{c c c c c c c c c c c }
    \hline \hline
    $\dot{M}_{inj} (10^{-8}M_\odot/\textrm{yr})$ &
    $\nu_{QPO}(Hz)$ & $\tilde T_{10}$ & $f_{rms} (\%)$ & $\Delta T_{sh}/t_{cr}$ & $\Delta\phi$ & $L_{bol}(erg/s)$ & $L_X(erg/s)$ & $E_{pk}(keV)$ & $\Gamma$ & $\eta$  \\ \hline
    
    0.5 & 2.50 & 80 & 10.0 & 1.20 & 0.63 & $5.9\times 10^{37}$ & $4.2\times10^{36}$ & 4600 & 0.9 & 0.12 \\ 
    
    1 & 2.53 & 70 & 9.1 & 1.13 & 0.58 &$ 7.2\times 10^{37}$ & $8.7\times 10^{36}$ & 2300 & 0.9 & 0.07 \\ 
    
    2 & 2.57 & 30 & 6.3 & 1.03 & 0.54 &$ 9.9\times 10^{37}$ & $2.1\times 10^{37} $& 1800 & 0.9 & 0.05 \\ 
    
    4 & 2.28 & 35 & 8.4 & 0.90 & 0.43 & $1.1\times 10^{38}$ & $3.3\times 10^{37}$& 580 & 1.0 & 0.03 \\  
    
    8 & 1.95 & 170 & 16.5 & 0.80& 0.32 & $1.4\times 10^{38}$ & $5.8\times 10^{37}$& 360 & 1.1 & 0.02 \\ 
    
    16 & 1.95 &1500 & 23.5 & -0.04 & -0.02 &$ 1.7\times 10^{38} $& $8.5\times 10^{37} $& 280 & 1.2 & 0.011\\ 
    
    32 & 2.12 & 2000 & 23.1 & -0.27 & -0.12 & $1.9\times 10^{38} $&$ 1.1\times 10^{38}$ & 230 & 1.5 & 0.006 \\ 
    
    64 & 2.38 & 50 & 5.3 & -0.5 & -0.27& $2.0\times 10^{38} $& $1.4\times 10^{38}$ & 150& 1.7 & 0.003 \\ \hline
  \end{tabular}
  \tablefoot{
  Table showing the various parameters of the problem as a function of the mass accretion rate $\dot{M}_{inj}$. For the definition of these parameters see the caption of Tab.~\ref{tab:3} and text. 
Other initial parameters are $R_c=10^9~\textrm{cm},~\tilde t_{en}=2.5,~t_{esc}/ t_{en}=10^3,~\alpha=0.1$ and $T_{eff}=4\times 10^6~\textrm{K}$.  
  }
\label{tab:4}
\end{table*}

\begin{figure}
\centering
\includegraphics[width = 0.5\textwidth]{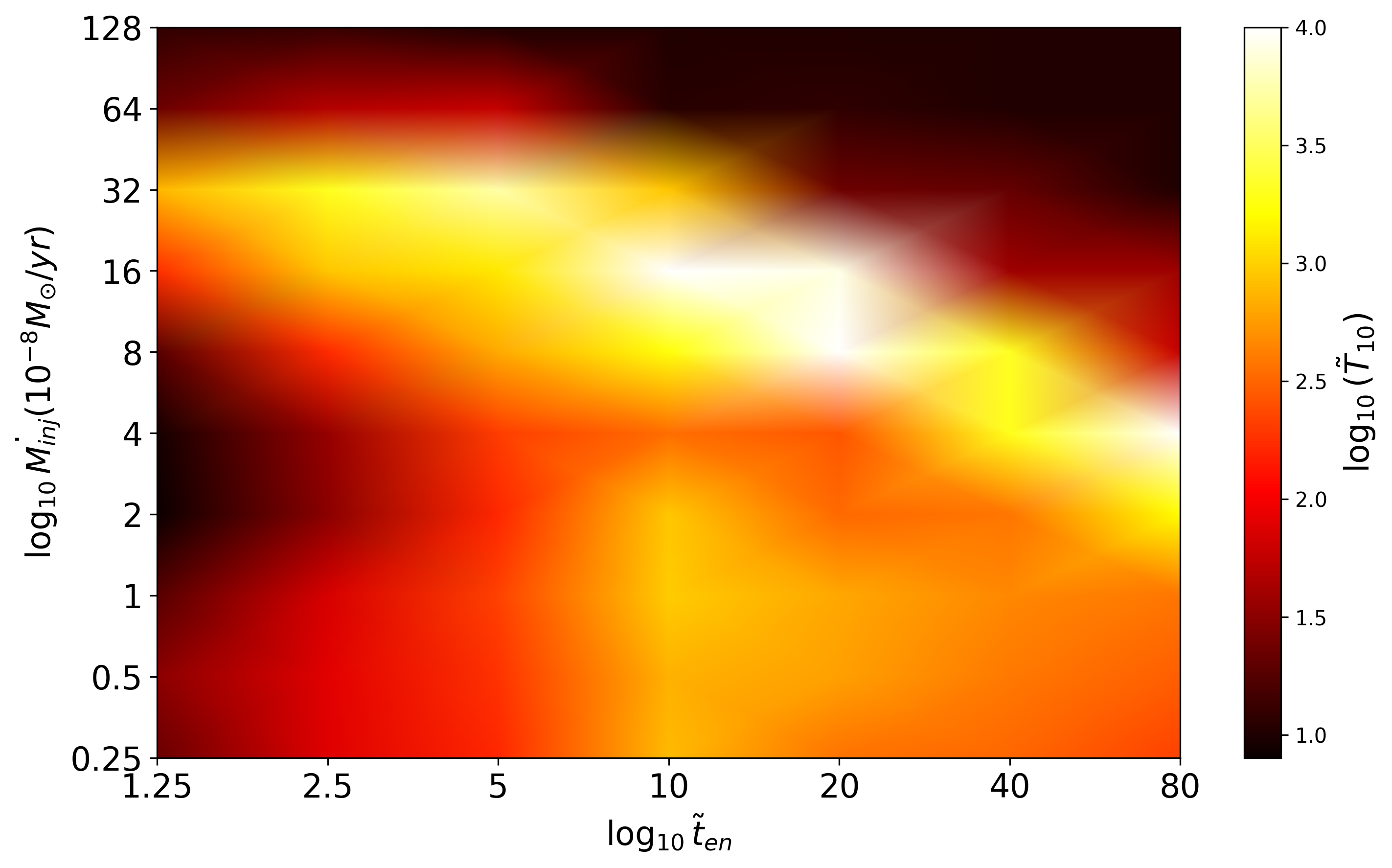}
\caption{Heat map of the parameter $\tilde T_{10}$ as a function of $\tilde t_{en}$ and $\dot{M}_{inj}$ for $R_c=10^9~\textrm{cm}$. Other parameters are $\alpha=0.1$ and $T_{eff}=4\times 10^6~\textrm{K}$. Dark regions correspond to light curves with strong decay.
} 
\label{figT10}
\end{figure}

Our numerical study reveals that the only non-monotonous  quantities  are the degree of damping, which we have expressed in terms of the quantity $\tilde T_{10}$, and the fractional rms $f_{rms}$, which, however, is related to the former, in the sense that heavily damped oscillations have small values of both $\tilde T_{10}$ and $f_{rms}$ and vice versa. Figure~\ref{figT10}
shows  a heat map for the $\tilde T_{10}$ parameter as a function of both $\tilde t_{en}$ and $\dot{M}_{inj}$. Dark regions indicate light curves showing strong damping. One can deduce that only certain combinations of (i) slow $\tilde t_{en}$ and high $\dot{M}_{inj}$
(the dark patch in the upper right side) and (ii) fast 
$\tilde t_{en}$ and intermediate $\dot{M}_{inj}$ produce damped oscillations. In addition, all runs with $\dot{M}_{inj}>3.2\times 10^{-7}M_\odot/\textrm{yr}$ and $\tilde t_{en}<1.25$ produce heavily damped oscillations. 
At any rate, judging from Fig.~\ref{figT10} one deduces that the transitions from heavily damped to lightly damped limit cycles do not follow some easily explainable pattern. 

\begin{figure}
\centering
\includegraphics[width = 0.35\textwidth]{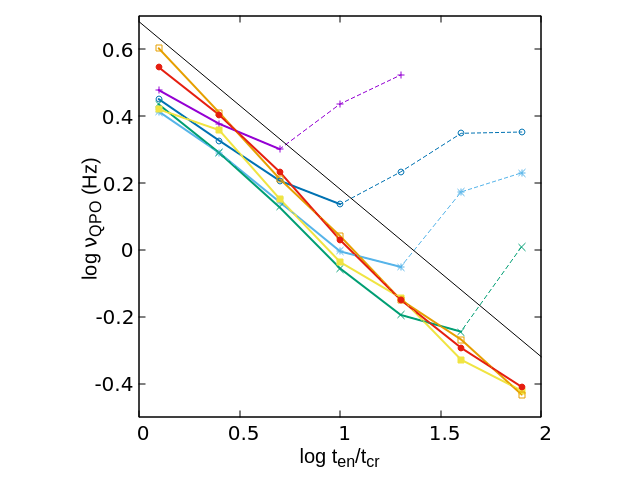}
\caption{Plot of the QPO frequency $\nu_{QPO}$ versus  the energization timescale $\tilde t_{en}$ for various values of $\dot{M}_{inj}$. 
Values of $\dot{M}_{inj}$, in units of 
$10^{-8}M_\odot/\textrm{yr}$, are $64$ (violet), $32$ (dark blue), $16$ (light blue), $8$ (green), $4$ (yellow), $2$ (orange) and $1$ (red). Other parameters are $R_c=10^9~\textrm{cm}$,  $\alpha=0.1$ and $T_{eff}=4\times 10^6~\textrm{K}$. The black line indicates the relation $\nu_{QPO}\propto t_{en}^{-1/2}$ given by Eq.~21.
}
\label{fignuten}
\end{figure}

Figure~\ref{fignuten} gives a plot of $\nu_{QPO}$ as a function of $\tilde t_{en}$ for various values of $\dot{M}_{inj}$.  The dashed lines connect those points that are characterized by strong damping ($\tilde T_{10}<50)$.
One notices the uniformity of $\nu_{QPO}$ if they are to exclude the dashed lines. Even if $\dot{M}_{inj}$ varies by almost two orders of magnitude, $\nu_{QPO}$ varies in most cases by less than a factor of 1.5.  
Therefore, the QPO frequency depends very weakly on $\dot {M}_{inj}$ -- note, however, that  $\dot {M}_{inj}$, as it was defined in the present paper, is not related directly to the X-ray luminosity but it is related, instead, to the electron injection at low energies, see Eq.~\ref{eq:qinj}. It is also interesting to note that roughly $\nu_{QPO}\propto \tilde{t}_{en}^{-1/2}$  -- compare the slope of the black line to that of the $\nu_{QPO}$ vs. $\tilde t_{en}$ curves. 

\begin{figure}
\centering
\includegraphics[width = 0.35\textwidth]{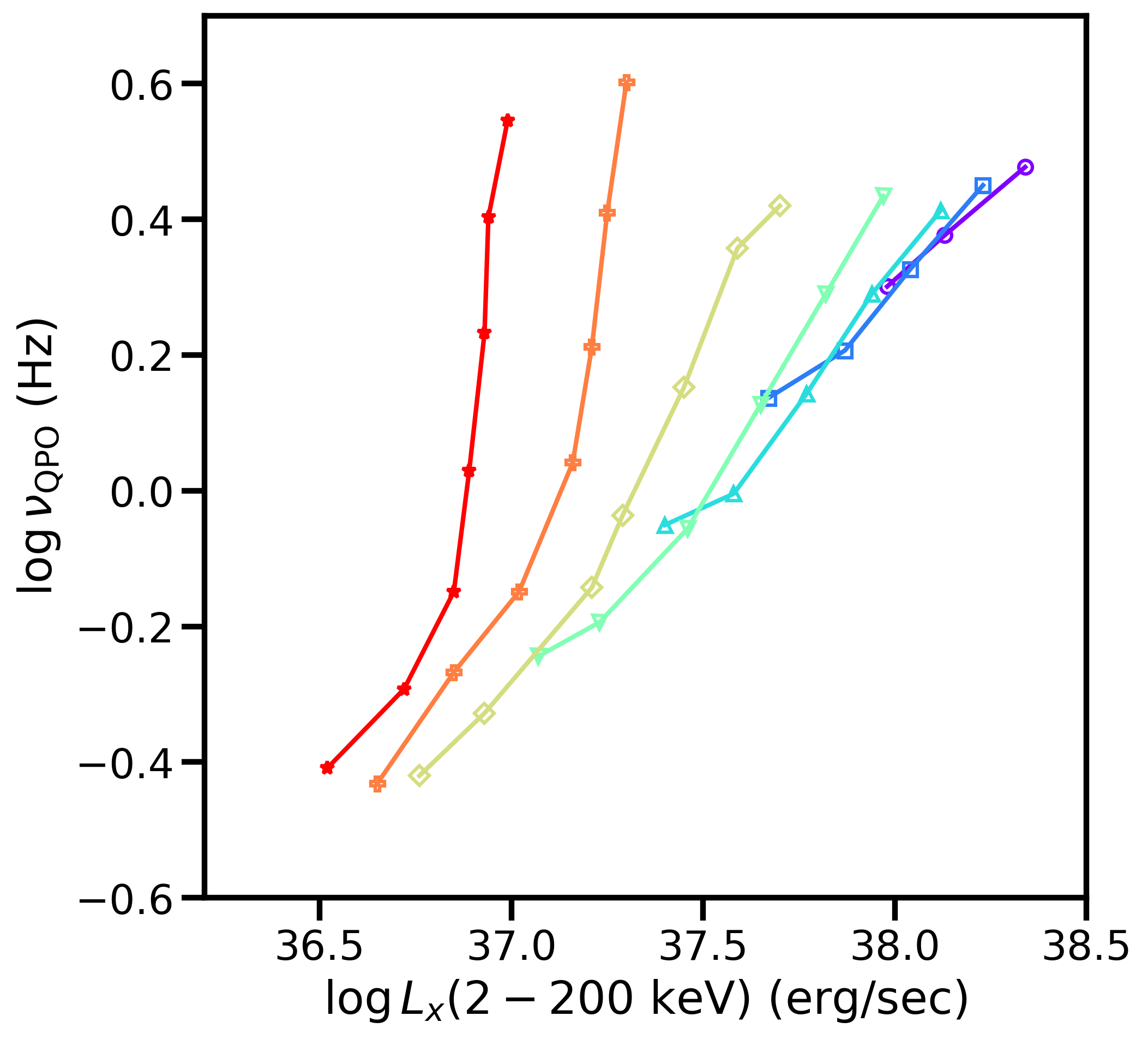}
\caption{Plot of the QPO frequency $\nu_{QPO}$ versus the $2-200~\textrm{keV}$ luminosity $L_{X}$ for various values of $\dot{M}_{inj}$. 
Values of $\dot{M}_{inj}$, in units of 
$10^{-8}M_\odot/\textrm{yr}$, are $32$ (violet), $16$ (dark blue), $8$ (light blue), $4$ (green), $2$ (yellow), $1$ (orange) and $0.5$. Other parameters are $R_c=10^9~\textrm{cm}$,  $\alpha=0.1$ and $T_{eff}=4\times 10^6~\textrm{K}$.
} 
\label{fignuLx}
\end{figure}

\begin{figure}
\centering
\includegraphics[width = 0.5\textwidth]{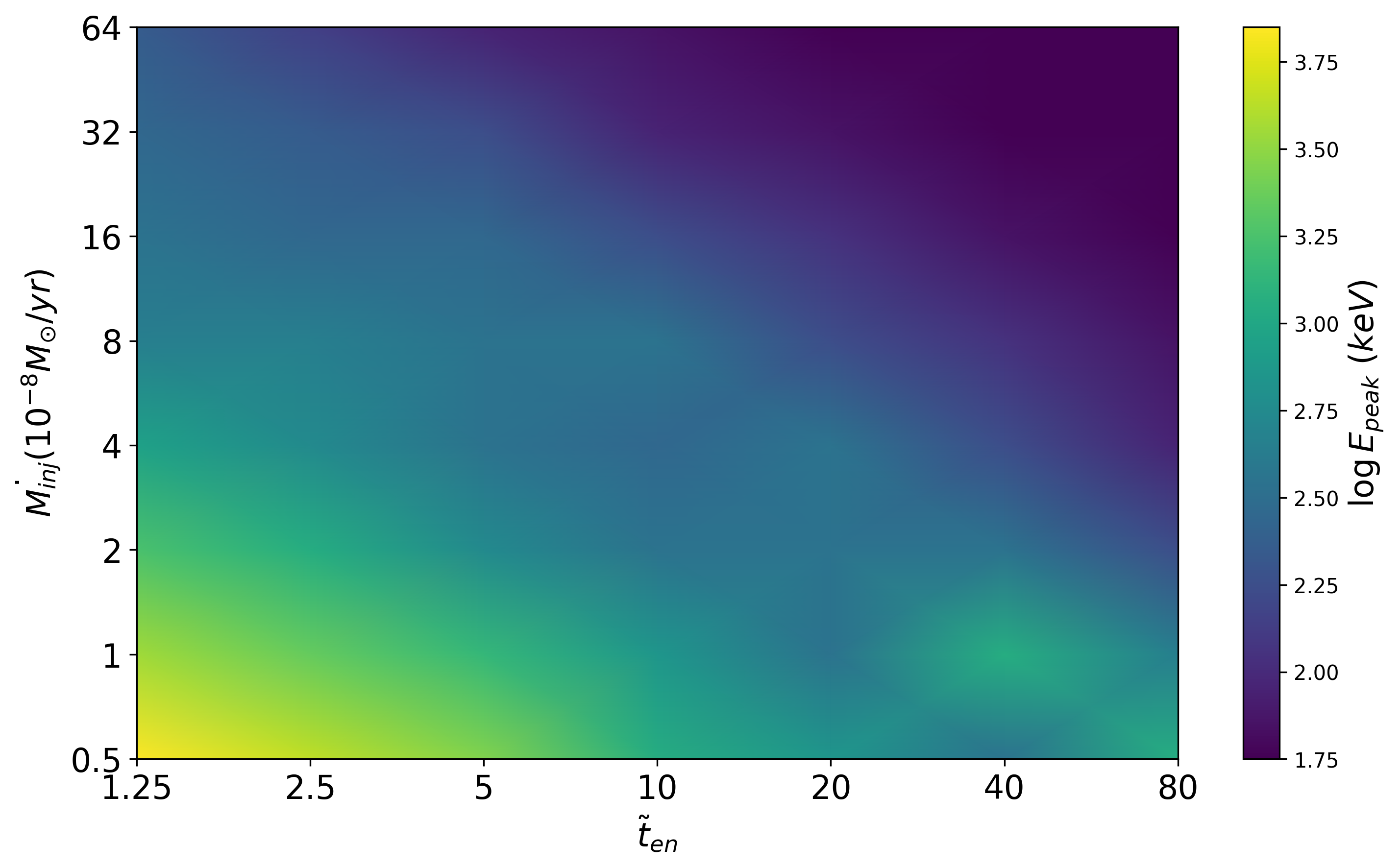}
\caption{Heat map of the average peak energy of the photon distribution $E_{pk}$ as a function of $\tilde t_{en}$ and $\dot{M}_{inj}$ for $R_c=10^9~\textrm{cm}$. Other parameters are $\alpha=0.1$ and $T_{eff}=4\times 10^6K$. Other parameters are $\alpha=0.1$ and $T_{eff}=4\times 10^6~\textrm{K}$. Dark blue regions indicate regimes with low values of $E_{pk}$. 
} 
\label{fightEpk}
\end{figure}

Figure~\ref{fignuLx} shows the relation of two observables, $L_X$ vs. $\nu_{QPO}$ for various values of $\dot{M}_{inj}$ and $R=10^9~\textrm{cm}$
-- here the points characterized by strong damping (connected with dashed lines in the previous figure) have been omitted for the  sake of clarity. The tendency is that cases with higher values of 
$\dot{M}_{inj}$ to reach higher luminosities, however the QPO frequencies cannot exceed values of around 3 irrespective of $\dot M_{inj}$. Note that the high $\dot{M}_{inj}$ cases cannot reach low $\nu_{QPO}$ values due to damping.
\begin{figure}
\centering
\includegraphics[width = 0.5\textwidth]{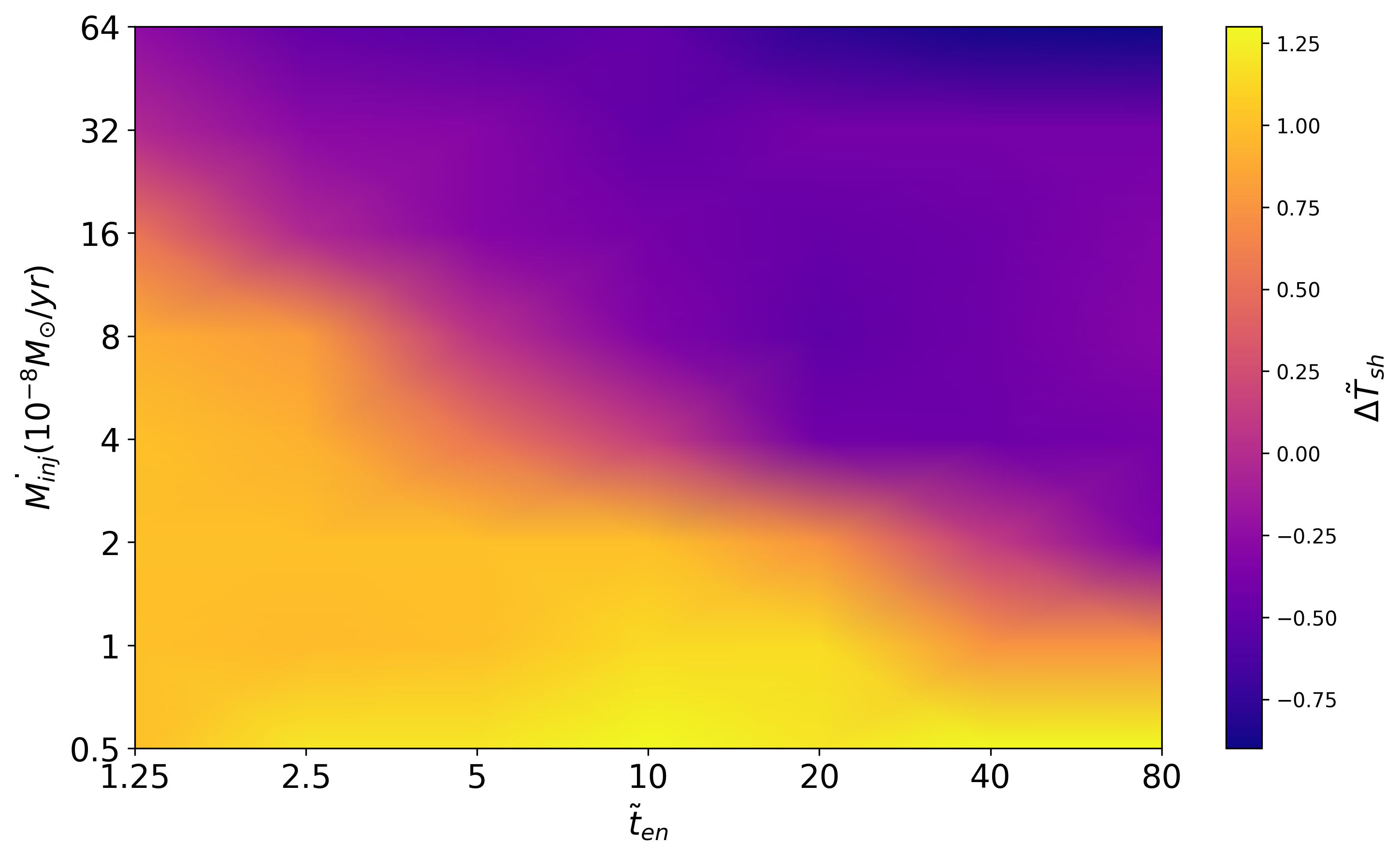}
\caption{Heat map of the time lag $\Delta T_{sh}$ (in units of $t_{cr}$)  as a function of $\tilde t_{en}$ and $\dot{M}_{inj}$ for $R_c=10^9~\textrm{cm}$. Other parameters are $\alpha=0.1$ and $T_{eff}=4\times 10^6K$. Violet color indicates negative lags, i.e. hard photons come before the soft ones in the prescribed energy bands.
} 
\label{fightDTsh}
\end{figure}

As an example of the variations of an output parameter in the $\tilde t_{en}-\dot{M}_{inj}$ plane, we depict in Figure~\ref{fightEpk} a heat map of $E_{pk}$. There is a tendency for this parameter to increase as $\tilde t_{en}$ and $\dot{M}_{inj}$ decrease. There are cases where $E_{pk}$ can enter the MeV regime (light green and yellow patches), however this is a sensitive function of the reprocessing parameter $\alpha$. Higher values of $\alpha$ tend to reduce drastically $E_{pk}$ -- see Sect.~\ref{sec:5.4}.

Figure~\ref{fightDTsh} gives another heat map, this time of the time lag $\Delta T_{sh}$   as a function of $\tilde t_{en}$ and $\dot{M}_{inj}$ for $R_c=10^9~\textrm{cm}$. In Appendix~\ref{app2} we give a physical interpretation of these time lags, however, as a rule of thump, we can note that negative time lags are produced from electron cooling while positive ones arise due to contributions from the gradual dominance of the spectrum from photons of the reprocessed component. As we show in Sect.~\ref{sec:5.5}, $\Delta T_{sh}$ is a sensitive function of  $T_{eff}$.

Bolometric luminosities, on the other hand, tend to be larger at high
$\dot{M}_{inj}$ and low $\tilde t_{en}$, reaching a 
value of $4\times10^{38}~erg/sec$, which is
$\simeq 30\%$ of the Eddington luminosity for a $10~M_\odot$ black hole, for 
$\dot{M}_{inj}=10^{-8}M_\odot/\textrm{yr}$ and $\tilde t_{en}=1.25$;
they are minimum (of the order of $~6\times10^{36}~erg/sec$) at the opposite corner of the heat map diagrams, i.e. for low $\dot{M}_{inj}$ and high $\tilde t_{en}$. X-ray luminosities follow more or less the same trend but their ratio to the bolometric luminosities does not remains constant, as already explained.

\subsection{Corona radius $R_c$ variations}\label{sec:5.3}

The present model shows a very simple scaling of the various parameters with $R_c$, as shown in a simplified, analytical way in  Sect.~\ref{sec:6}. For this to work, one needs to scale $R_c$ and
$\dot{M}_{inj}$ by the same factor. 
Thus, if both $R_c$ and $\dot{M}_{inj}$ increase by some factor,  then $\nu_{QPO}$
decreases while 
$~L_{bol}$ and $L_X$ increase, all by the same factor. The rest of the output parameters remains unchanged.

This helps to use the results at $R_c=10^9~\textrm{cm}$ as template and readily derive the output parameters for other corona radii. Thus, following the same procedure as before, we show in Tab.~\ref{tab:5} the results one obtains if they are to keep $\tilde t_{en}$ and 
$\dot{M}_{inj}$ constant and vary $R_c$.
Here one notices that $\nu_{QPO}$ varies roughly as $R_c^{-1}$, there is a  gradient of $\Delta T_{sh}$ and $\Delta\phi$ from negative to positive values, while  both $L_{bol}$ and $L_X$ are increasing functions of $R_c$ ($L_{bol}$ varies almost linearly with $R_c$, while $L_X$ as $R_C^{1/2}$). Also $E_{pk}$ increases with $R_c$ and as a result the spectrum becomes flatter.

\begin{table*}[h]
  \centering
  \caption{Results as a function the corona radius $R_c$.}
  \begin{tabular}{c c c c c c c c c c c }
    \hline \hline
    $R_c(cm)$ &
    $\nu_{QPO}(Hz)$ & $\tilde T_{10}$ & $f_{rms} (\%)$ & $\Delta T_{sh}/t_{cr}$ & $\Delta\phi$ & $L_{bol}(erg/s)$ & $L_X(erg/s)$ & $E_{pk}(keV)$ & $\Gamma$ & $\eta$   \\ \hline
   $2.5\times 10^8$ & 9.64 & 50 & 4.4 & -0.5 & -0.26 & 4$.9\times 10^{37}$ & $3.6\times 10^{37} $& 180 & 1.7 & 0.003 \\ 
    
    $5\times 10^8$ & 4.24 & 2000 & 23.1 & -0.27 & -0.12 & $9.5\times 10^{37}$ & $5.5\times 10^{37}$ & 230 & 1.6 & 0.006 \\ 
    
   $ 1\times 10^9$ & 1.95 &1500 & 23.5 & -0.04 & -0.02 &$ 1.7\times 10^{38}$ & $8.5\times 10^{37} $& 280 & 1.2 & 0.011 \\ 
    
   $ 2\times10^9 $& 0.98 & 150 & 9.7 & 0.8 & 0.32 & $2.9\times 10^{38}$ & $1.2\times 10^{38}$&360  & 1.1 & 0.018 \\ 
    
   $ 4\times10^9$ & 0.56 & 35 & 5.7 & 0.93 & 0.43 & $4.6\times 10^{38} $& $1.3\times 10^{38}$& 580 & 1. & 0.03 \\ 
 \hline
   
  \end{tabular}
  \tablefoot{Table showing the various parameters of the problem as a function the corona radius $R_c$. For the definition of these parameters see the caption of Tab.~\ref{tab:3} and text. 
   Other initial parameters are $\tilde t_{en}=2.5,~\dot M_{inj}=16\times10^{-8}M_\odot/\textrm{yr},~t_{esc}/ t_{en}=10^3,~\alpha=0.1$ and  $T_{eff}=4\times 10^6~\textrm{K}$.
  }

\label{tab:5}
\end{table*}

\subsection{Reprocessing fraction $\alpha$ variations}\label{sec:5.4}
Another parameter that is unknown, but plays a role in QPO formation, is the reprocessing parameter $\alpha$. Table~\ref{tab:6} shows the output parameters of the problem in the case where $\alpha$ is varying. 

\begin{table*}[h]
  \centering
  \caption{Results as a function of the reprocessing fraction  $\alpha$.}
  \begin{tabular}{c c c  c c c c c c c c}
   \\\hline\hline
    $\alpha$&$\nu_{QPO}(Hz)$ & $\tilde T_{10}$ & $f_{rms} (\%)$ & $\Delta T_{sh}/t_{cr}$ & $\Delta\phi$ & $L_{bol}(erg/s)$ & $L_X(erg/s)$ & $E_{pk}(keV)$ & $\Gamma$ & $\eta$  \\ \hline
    
    0.05 & 1.76 & 900& 33.5 & -0.13 & -0.05& $2.8\times 10^{38}$ &$ 9.9\times 10^{37}$ &580 & 1.1 & 0.018 \\ 
    
    0.1 & 1.95 &1500 & 23.5 & -0.04 & -0.02 & $1.7\times 10^{38} $& $8.5\times 10^{37}$ & 280 & 1.2 & 0.011 \\ 
   
    0.2& 2.17 & 2000 & 18.5 & -0.04 & -0.02 &$ 1.2\times 10^{38} $& $8.\times 10^{37}$ & 150 & 1.5 & 0.095\\ 
    
   0.4& 2.47 & 30 & 13.8 & 0.0 & 0.0 & $8.5\times 10^{37}$ & $7.3\times 10^{37} $&  40 & 1.8 & 0.066\\ \hline

 \end{tabular}
  \tablefoot{Table showing the various parameters of the problem as a function of the reprocessing fraction  $\alpha$. For the definition of these parameters see the caption of Tab.~\ref{tab:3} and text.    Other initial parameters are $R_c=10^9~\textrm{cm},~\tilde t_{en}=2.5,~\dot M_{inj}=16\times10^{-8}M_\odot/\textrm{yr},~t_{esc}/ t_{en}=10^3,$ and $T_{eff}=4\times 10^6~\textrm{K}$. }
\label{tab:6}
\end{table*}
One notices that changes in $\alpha$ affect most of the output parameters, but only mildly -- i.e. despite the fact that $\alpha$ varies by a factor of 8, most parameters change by a factor of 2 to 3, while some, such as $\Delta T_{sh}$ remain practically unaffected. It is also worth mentioning that  even a small value of $\alpha$ is enough to set the system go through limit cycles. In this case, however, the electrons have to reach relatively high energies before they trigger the feedback from the reprocessed photons. As a result, the spectra become flatter for decreasing $\alpha$. Generally speaking, low values of $\alpha$ tend to produce flat spectra ($\Gamma<2$) and high $E_{pk}$, while the opposite holds when $\alpha$ is increasing. A final comment is that high values of $\alpha$ cause strong damping, however this depends on the other parameters of the problem as well.

\subsection{Reprocessing temperature $T_{eff}$ variations}\label{sec:5.5}

Another input parameter that has an impact on the spectro-temporal properties of the oscillations is the temperature $T_{eff}$ that characterizes the reprocessed emission of the accretion disk.Table~\ref{tab:7} tabulates the usual parameters for four values of $T_{eff}$. Here once again we notice that most of the output parameters show much slower variations than the input ones; however, we should mention that $\Delta T_{sh}$, and by extension $\Delta\phi$, show a rather radical change from negative to positive as $T_{eff}$ increases. This means that phase lags are a sensitive function of the choice of $T_{eff}$. This is a result of the exponential cutoff in the Wien part of the reprocessed component -- see  Appendix~\ref{app2}.

\begin{table*}[h]
  \centering
  \caption{Results as a function the reprocessed disk temperature $T_{eff}$.}
  \begin{tabular}{ c c c c c c c c c c c}
    \hline \hline
    $T_{eff}(K)$&    $\nu_{QPO}(Hz)$ & $\tilde T_{10}$ & $f_{rms} (\%)$ & $\Delta T_{sh}/t_{cr}$ & $\Delta\phi$ & $L_{bol}(erg/s)$ & $L_X(erg/s)$ & $E_{pk}(keV)$ & $\Gamma$ & $\eta$  \\ \hline
    
   $ 1\times 10^6$ & 2.14 & 200& 11.5 & -0.7 &-0.31  & $1.2\times10^{38} $&$ 3.0\times 10^{37}$ &280 & 1.9 & 0.008\\ 
    
   $ 2\times 10^6$ & 2.06 & 1000& 21.8 & -0.5 & -0.22 & $1.5\times 10^{38}$ & $7.8\times 10^{37}$ & 280 & 1.6 & 0.009\\ 

   $ 4\times 10^6$ & 1.95 &1500 & 23.5 & -0.04 & -0.02 & $1.7\times10^{38}$ & $8.5\times 10^{37} $& 280 & 1.2 & 0.011 \\ 
    
   $8\times 10^6$& 1.88 & 10000 & 28.2 & 0.8 & 0.3 & $2.2\times 10 ^{38} $& $9.8\times 10^{37} $& 580 & 1.1 & 0.014 \\ 
\hline
  \end{tabular}
 \tablefoot{Table showing the various output parameters of the problem as a function the reprocessed disk temperature $T_{eff}$. For the definition of these parameters see the caption of Tab.~\ref{tab:3} and text. 
   Other initial parameters are $R_c=10^9~\textrm{cm},~\tilde t_{en}=2.5,~\dot M_{inj}=16\times10^{-8}M_\odot/\textrm{yr},~t_{esc}/ t_{en}=10^3,~\alpha=0.1$.  }
\label{tab:7}
 \end{table*}

\subsection{Ratio of escape to energization timescales  $t_{esc}/t_{en}$ variations}\label{sec:5.6}

The last input parameter that has an impact on the spectro-temporal properties of the oscillations is the ratio of the escape to energization timescales. As it has already been mentioned in Sect.~\ref{sec:2}, as this ratio increases, the slope of the electron distribution becomes flatter tending asymptotically to $-1$. Energy-wise this translates to more electron energy being transferred to the high end making the feedback stronger.  Table~\ref{tab:8} shows that for $t_{esc}=t_{en}$ the oscillations are damped, however for $t_{esc}/t_{en}>10$ damping becomes negligible.

\begin{table*}[h]
  \centering
  \caption{Results as a function the ratio $t_{esc}/t_{en}$.}
  \begin{tabular}{ c c c c c c c c c c c}
    \hline \hline
    $t_{esc}/t_{en}$&    $\nu_{QPO}(Hz)$ & $\tilde T_{10}$ & $f_{rms} (\%)$ & $\Delta T_{sh}/t_{cr}$ & $\Delta\phi$ & $L_{bol}(erg/s)$ & $L_X(erg/s)$ & $E_{pk}(keV)$ & $\Gamma$ & $\eta$  \\ \hline
    
   $1$ & 1.92 & 35& 11.5 & 0.63 &0.25  & $2.1\times10^{37} $&$ 2.1\times 10^{36}$ &2800 & 1.3 & 0.003\\ 

$ 10$ & 1.89 & 1000 & 21.8 & 0.27 & 0.11 & $1.5\times 10^{38}$ & $7.9\times 10^{37}$ & 280 & 1.2 & 0.010\\

   $100$& 1.95 & 1500 & 23.3 & -0.01 & -0.01 & $1.7\times 10 ^{38} $& $8.4\times 10^{37} $& 280 & 1.2 & 0.011 \\

   $ 1000$ & 1.95 &1500 & 23.5 & -0.04 & -0.02 & $1.7\times10^{38}$ & $8.5\times 10^{37} $& 280 & 1.2 & 0.011 \\ 

\hline
   
  \end{tabular}
 \tablefoot{Table showing the various output parameters of the problem as a function the ratio $t_{esc}/t_{en}$.
 For the definition of these parameters see the caption of Tab.~\ref{tab:3} and text. 
   Other initial parameters are $R_c=10^9~\textrm{cm},~\tilde t_{en}=2.5,~\dot M_{inj}=16\times10^{-8}M_\odot/\textrm{yr},~\alpha=0.1$ and $T_{eff}=4\times 10^6~\textrm{K}$.  }
\label{tab:8}
 \end{table*}

\section{A simplified analysis of the electron-photon system}\label{sec:6}

In this section we will show in a simplified manner the basic principles behind the natural frequency occurrence discussed in the main body of the paper. As in MPK, we will assume that the electrons and photons do not depend explicitly on energy, i.e. we will treat energy densities instead of differential densities. Then the electron and photon equations can be written
\begin{equation}
\frac{\partial u_e}{\partial t}  =\frac{u_e}{t_{en}} 
- \frac{u_e}{t_{loss}} 
\label{eq:kinetic1}
\end{equation}
and 
\begin{equation}
\frac{\partial u_\gamma}{\partial t}  + \frac{u_\gamma}{t_{cr}} 
= Q_{sources}.
\label{eq:kinetic2}
\end{equation}
Here $t_{loss}$ is the characteristic time scale for electron losses while $Q_{sources}$ denotes the photon energy gains. 

We next assume that there are two types of processes: (i) linear, like synchrotron radiation, ICS on disk photons, etc and (ii) non-linear like ICS on reprocessed photons, see Eq.~(\ref{equation7}) and (\ref{equation8}). Thus we write
for the electron equation
\begin{equation}
\frac{\partial u_e}{\partial t}  = \frac{u_e}{t_{en}} 
-\frac{u_e}{t_{lin}}-\frac{u_e}{t_{non,lin}}
\label{eq:kinetic3}
\end{equation}
with ${t_{lin}}$ and $t_{non,lin}$ the two loss timescales corresponding to the linear and non-linear processes, respectively.
Similarly, we can write for the photon equation
\begin{equation}
\frac{\partial u_\gamma}{\partial t} + \frac{u_\gamma}{t_{cr}} 
=  Q_{lin}+Q_{non,lin}.
\label{eq:kinetic4}
\end{equation}
where, due to energy conservation, each photon rate ($Q_{lin}$ and $Q_{non,lin}$) should be equal to the corresponding loss term entering the electron equation. We proceed next to examine this system of equations according to the combinations of interest to the present paper.\\

\subsection{Purely non-linear case}

We begin by
ignoring, for the moment, the linear terms. We  set 
$t^{-1}_{non,lin}\simeq \sigma_Tc(u_s/m_ec^2)$ where $u_s=\alpha u_\gamma$ the reprocessed soft energy density 
and $\alpha$ the reprocessing parameter. We normalize  the energy densities by using $\tilde u_{e,\gamma}=\sigma_TR_c(u_{e,\gamma}/{m_ec^2})$ and time by 
using $\tilde t=t/t_{cr}$. Then the set of Eqs.~(\ref{eq:kinetic3}) and (\ref{eq:kinetic4})  becomes 
\begin{equation}
\frac{\partial \tilde u_e}{\partial \tilde t}  = \frac{\tilde u_e}{\tilde t_{en}} 
-  \alpha\tilde u_\gamma \tilde u_e 
\end{equation}
and 
\begin{equation}
\frac{\partial \tilde u_\gamma}{\partial \tilde t}=  - \tilde u_\gamma +
\alpha \tilde u_\gamma \tilde u_e
\end{equation}
This set of equations is the classic predator-prey (Lotka-Volterra) system.
As it is known from theory, e.g. see \cite{Strogatz}, this system shows undamped oscillations with a natural frequency that is given, for the present case, by 
$\tilde \nu_0=\tilde t_{en}^{-1/2}/2\pi$
or, introducing units, 
\begin{equation}
\nu_0=(c/2\pi R_c)\tilde t_{en}^{-1/2}.
\label{eq:nuanal}
\end{equation}
This is depicted with a black line in Fig.~\ref{fignuten}. While it
matches very well the numerically derived slopes, it overestimates the calculated $\nu_{QPO}$ by $\sim 50\%$.
Furthermore, since the equilibrium points are $\tilde u_{e,eq}=\alpha$ and
$\tilde u_{\gamma,eq}=\alpha\tilde t_{en}$, and since, by definition, $\ell_\gamma\equiv \tilde u_\gamma$, it becomes evident that the compactness $\ell$ is also scale-free. 
From Eq.~\ref{eq:nuanal} and the arguments given above,
it becomes evident from that the  natural frequency scales as $R_c^{-1}$ and the luminosity as $R_c$ -- see 
Section 5.3. Furthermore, the found dependence of $\tilde u_{\gamma,eq}$ on the reprocessing parameter explains, at least qualitatively, the calculated $L_{bol}$ behavior with $\alpha$. 

\subsection{Introducing damping: Synchrotron radiation}

Now the system of
Eqs.~(\ref{eq:kinetic3}) and (\ref{eq:kinetic4}), in addition to the non-linear term examined above, contains a linear term of the form $u_e/t_{syn}$ 
where $t_{syn}=(\sigma_Tcu_B/m_ec^2)^{-1}$ with $u_B$ the magnetic field energy density.  This term transfers energy from electrons to photons at a constant rate and causes damping in the system which now has the form 
\begin{equation}
\frac{\partial \tilde u_e}{\partial \tilde t}  = \frac{\tilde u_e}{\tilde t_{en}} 
-  \alpha\tilde u_\gamma \tilde u_e -d_B\tilde u_e
\label{eq:synel}
\end{equation}
and 
\begin{equation}
\frac{\partial \tilde u_\gamma}{\partial \tilde t}=  - \tilde u_\gamma +
\alpha \tilde u_\gamma \tilde u_e +d_B\tilde u_e,
\label{eq:synph}
\end{equation}
where $d_B=\sigma_TR_c(u_B/m_ec^2)$ is the magnetic compactness.
From standard non-linear theory we obtain that 
the natural frequency becomes in this case
\begin{equation}
\tilde \nu_0=\frac{1}{2\pi}\left[(\frac{1}{\tilde t_{en}}-d_B)-\frac{1}{4}(\tilde t_{en}d_B)^2\right]^{1/2}    
\end{equation}
provided that the expression in the parenthesis is greater than 0. The latter  gives the condition required for oscillations to occur:
\begin{equation}
d_B< d_{B,crit}\equiv \frac{2}{\tilde t^2_{en}}[(1+\tilde t_{en})^{1/2}-1].    
\end{equation}
From here we can obtain a limiting value of $B$ 
\begin{equation}
B_{cr}=1.7\times10^5R_9^{-1/2}{\tilde t^{-1}_{en}}\left[(1+\tilde t_{en})^{1/2}-1\right]^{1/2}~G
\label{eq:Bmx}
\end{equation}
above which the system becomes overdamped. We also note that since the damping rate is $\lambda=\frac{1}{2}d_B\tilde t_{en}$, small values of $d_B$ (for given $\tilde t_{en}$) produce negligible damping and this is the underlying assumption of the current paper where synchrotron has been neglected. Damping increases with $d_B$, until it reaches the limiting value $d_{B,crit}$ above which oscillations are quenched. A numerical treatment of the effects of synchrotron radiation is given in Appendix~\ref{app3}.

\subsection {Introducing damping: ICS on disk photons}

Assuming that in addition to the reprocessed photons, the disk radiates its own photons which have energy density $U_{ds}$ (see Sect.~\ref{sec:3}), the system becomes identical to Eqs.~\ref{eq:synel} and \ref {eq:synph} with $\tilde u_{ds}$ replacing $d_B$. 
For $t_{en}=2.5$ one gets $\tilde u_{ds,mx}=0.3$ which can be readily compared to the case shown in Fig.~\ref{f_light} where damping becomes significant for $\ell_{ds}\equiv\tilde u_{ds}\simeq 0.1$.

\section{The case of GRS 1915+105}\label{sec:7}

Although it would have been tempting to use the method outlined in Sect.~\ref{sec:5} and try to produce detailed $\chi^2$ fits to the observational data, the number of free parameters would not have offered a physical insight into the underlying mechanism of the source. In contrast, it would be more interesting to see whether a systematic  change in some fundamental quantity of our model would produce results that could be compared favorably to observational data, because this could give us a hint of the physics of the source.
Therefore, we have tried to compare data to trends produced in our model by changing systematically one of the free parameters. 
As an example, we take the case of GRS 1915+105, that is a well-studied source, see, e.g., 
\cite{Morganetal97, Trudolyubovetal99,Reigetal00,Quetal10}. 
Recently, \cite{Zhangetal20} analyzed 620 RXTE observations of the source and found that the QPO phase lag decreases with QPO frequency 
and change sign from positive
to negative at a frequency around 2 Hz.

\begin{figure}
\centering
\includegraphics[width = 0.42\textwidth]{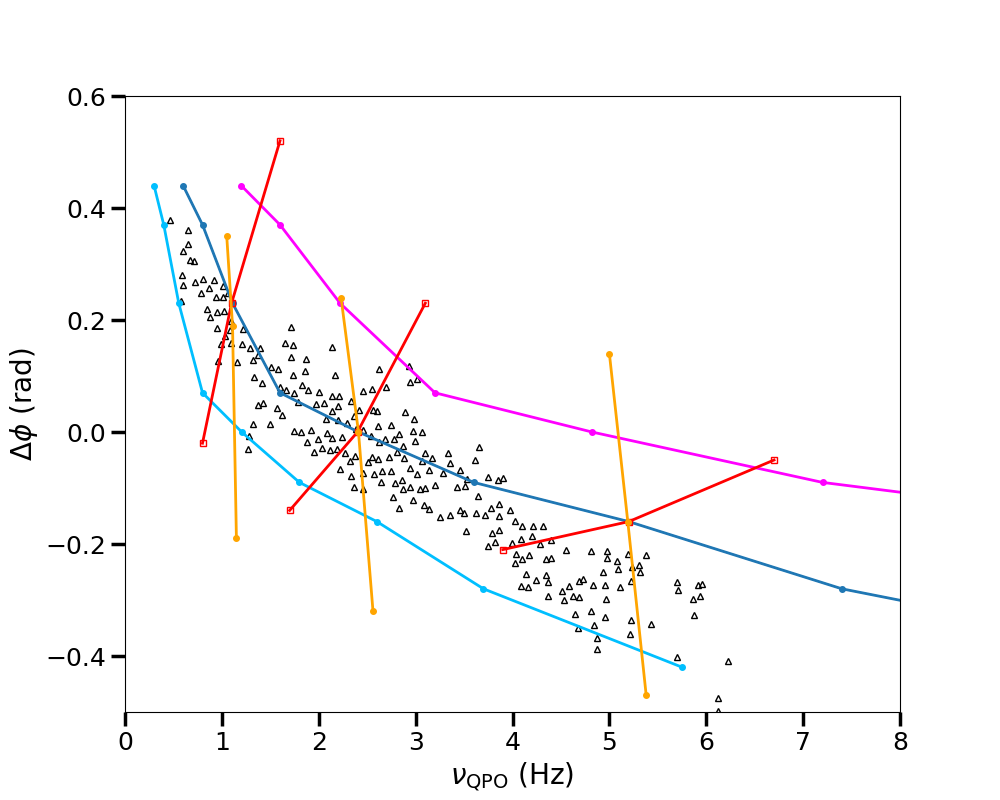}
\caption{Plot of the 
phase lag $\Delta\phi$ versus the
QPO frequency $\nu_{QPO}$.
The violet line corresponds to 
$\dot{M}_{inj}=8\times
10^{-8}M_\odot/\textrm{yr}$ for $R_c$ between $10^{9.3}~\textrm{cm}$ and 
$R_c=10^{7.95}~\textrm{cm}$,
the dark blue line corresponds to 
$\dot{M}_{inj}=16\times
10^{-8}M_\odot/\textrm{yr}$ for $R_c$  between $10^{9.6}~\textrm{cm}$ and
$10^{8.25}~\textrm{cm}$,
while the light blue line corresponds to 
$\dot{M}_{inj}=32\times
10^{-8}M_\odot/\textrm{yr}$ for $R_c$  between $10^{9.9}~\textrm{cm}$ to 
$10^{8.85}~\textrm{cm}$. In all cases the radius $R_c$ is decreasing from left to right.
The other parameters are $\tilde t_{en}=2.5$,  $\alpha=0.4$ and $T_{eff}=4\times 10^6~\textrm{K}$.
The red lines show the results of varying $\tilde t_{en}$ by a factor of 2 (i.e. between .25 and 5) when $\dot{M}_{inj}=16\times
10^{-8}M_\odot/\textrm{yr}$ and the corona radius is  $R_c=10^{9.3}~\textrm{cm},~10^{9}~\textrm{cm},$
and $10^{8.7}~\textrm{cm}$ (l to r). As $\tilde t_{en}$ decreases, the curves move up and to the right.
Orange lines repeat the same procedure with $T_{eff}$ varying by a factor of 2 (i.e. between $2$ and $8\times 10^6~\textrm{K}$). As $T_{eff}$ increases, the curves move upwards. The data points are taken from  \cite{Zhangetal20}
.}
\label{figGRS1}
\end{figure}

The simplest approach 
to compare our model
to the above results
is to consider changes in only one of our parameters. Since the observations indicate that $\nu_{QPO}$ changes by about one order of magnitude, this could mean that it could be achieved either with variations in $\tilde t_{en}$ or in $R_c$. The phase lag, however, changes from negative to positive by almost $1~ rad$ and a brief inspection on Tab.~\ref{tab:3} indicates that this cannot be achieved by changes in $\tilde t_{en}$ alone. Therefore, we consider changes in $R_c$ and the results are shown in Fig.~\ref{figGRS1}. 
This shows the data from  
\cite{Zhangetal20}
(black triangular points) along with various lines produced with the present model. 
\begin{figure}
\centering
\includegraphics[width = 0.37\textwidth]{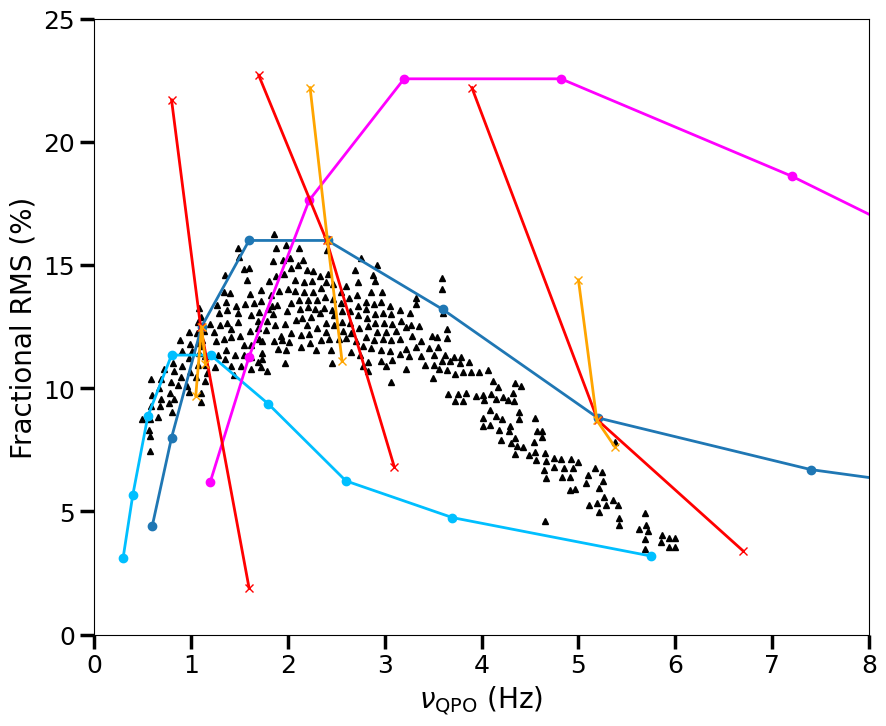}
\caption{Plot of the fractional RMS versus the
QPO frequency $\nu_{QPO}$ for the cases depicted in Fig.~\ref{fightEpk}. The fractional RMS increases with decreasing $\tilde t_{en}$ (red lines) and increasing $T_{eff}$ (orange lines). The data points are taken from  \cite{Zhangetal20}
} 
\label{figGRS2}
\end{figure}
The violet, dark and light blue lines represent results obtained for  $\dot M=8,~16$ and $32\times10^{-8}M_\odot/\textrm{yr}$, respectively keeping all other parameters constant but varying $R_c$. For instance, 
 the dark blue line curve shows results obtained by changing the  corona radius from $R_c=10^{9.6}~\textrm{cm}$ (upper left side) to $R_c=10^{8.55}~\textrm{cm}$ (lower right side) using a logarithmic step of $0.15$ between consecutive runs.
 Having obtained this particular curve, the others can be scaled according to the findings of Sect.~\ref{sec:5.3}. Therefore, the pair $(\nu_{QPO},\Delta\phi)$ at $R_c=10^{9.6}~\textrm{cm}$ for $\dot M=16\times10^{-8}M_\odot/\textrm{yr}$ will be mapped at $(\nu_{QPO}/2,\Delta\phi)$ 
  for $R_c=10^{9.9}~\textrm{cm}$ when $\dot M=32\times10^{-8}M_\odot/\textrm{yr}$. Similarly, it will be mapped 
at $(2\nu_{QPO},\Delta\phi)$
   for $R_c=10^{9.6}~\textrm{cm}$ when $\dot M=16\times10^{-8}M_\odot/\textrm{yr}$.

Having obtained a general trend that shows an overall similarity to the observations, a next step is to investigate how the system responds to changes in the other parameters. For this we have chosen three 
radii ($R_c=10^{9.3}~\textrm{cm},~10^9~\textrm{cm}$ and $10^{8.7}$~\textrm{cm}) for the case  $\dot M=16\times10^{-8}M_\odot/\textrm{yr}$ and increased/decreased $\tilde t_{en}$ by a factor of 2 while keeping all the other parameters fixed. The results are depicted with red lines in Fig.~\ref{figGRS1}. The red lines moving up and to the right indicate a decrease in $\tilde t_{en}$ -- see Tab.~\ref{tab:3}. 

We can repeat the same procedure for showing changes in the $\Delta\phi-\nu_{QPO}$ plane caused by variations in $T_{eff}$. So we performed runs increasing/decreasing $T_{eff}$ by a factor of 2 while keeping all other parameters fixed. The results are shown with an orange line in Fig.~\ref{figGRS1}.
In all three cases shown, $\Delta\phi$ increases with increasing $T_{eff}$. Furthermore, we note that the orange lines are almost vertical, i.e. temperature changes, at least in the range specified here, do not produce significant changes in $\nu_{QPO}$; however, they have a significant effect on $\Delta\phi$. This means that the present model is rather sensitive on the reprocessed disk temperature, a fact already mentioned in Sect.~\ref{sec:5.5}. 

%


Most of the observational points lie between the lines of $\dot M=16$ and $32\times10^{-8}M_\odot/\textrm{yr}$. 
Alternatively, one could assume that the spread of the data is caused by a constant accretion rate and fluctuations of a factor of 2 in either the energization timescale or the reprocessed disk temperature. 

Figure~\ref{figGRS2} shows the calculated fractional RMS vs. $\nu_{QPO}$ (colored lines) plotted over the GRS 1915+105 data (black symbols). The colors of lines correspond  to the cases examined in Fig.~\ref{figGRS1}. Once again most points lie between the dark and light blue lines, i.e. 
 $\dot M=16$ and $32\times10^{-8}M_\odot/\textrm{yr}$. However, we caution the reader that the fractional RMS was calculated from the light curves using a python routine;
 also, the shown value should be considered as an upper limit because the RMS was calculated using only a few cycles of the light curves. Therefore, the results shown here are only indicative.

Clearly, one could expand the approach taken above, by varying two or more of the input parameters. For instance, they could argue that by increasing $\dot M_{inj}$, $R_c$ decreases and then combine the results shown in the corresponding Tables to obtain relations between the various observables. This, however, requires some further physical justification and it it lies outside the aims of the present paper. 

Finally, it would be interesting to compare our results with the ones of \cite{Karpouzasetal21}
who, however, used a different approach 
to the  time-dependent Comptonization model
with feedback. Both models require large corona radii (of the order of $~10^3R_g$), assuming a black hole mass of $12.4~M_\odot$ \citep{Reidetal14} especially when $\nu_{QPO}$ is small. Furthermore, they both find that $\nu_{QPO}$ increases as $R_c$ decreases. However, in the \cite{Karpouzasetal21} model, $R_c$ starts increasing again as $\nu_{QPO}$ increases past $2~Hz$, while in our case $R_c$ keeps monotonically decreasing. 
Nevertheless, it should be mentioned that in \cite{Karpouzasetal21}
emphasis was given in fitting  simultaneously the phase lag/RMS data, so all parameters changed in order to obtain the best $\chi^2$; on the other hand, in our case the emphasis was given on the trends obtained by changing only $R_c$ while keeping the rest of the parameters fixed. 


\section{Summary-Discussion}\label{sec:8}
Aim of the present paper was to verify and extend the results of \citetalias{MPK22}, i.e. to show that QPOs arise naturally in a disk-corona system. 
For this we applied a one-zone model for the evolution of electrons and photons inside the corona  and solved the coupled kinetic equations for the two species distribution functions. We  assumed that the electrons were energized by some unspecified mechanism 
and that they 
were losing energy by ICS on soft photons. 
The origin of these photons was assumed to come from the reprocessing of the  electrons inverse Compton radiation on the disk.  This simple prescription is adequate to make the system go through limit-cycle behavior. As shown in Appendix~\ref{app1},  as  electrons gain energy, their ICS radiation increases and so does the reprocessed radiation. At some point the latter has increased beyond a critical point and it is able to overtake energization, cooling thus efficiently  the electrons. 
This leads to a reduction of both  the ICS radiation and  its reprocessed radiation, therefore the electrons are able once again to gain energy  and the cycle repeats itself. For this to occur one needs the disk to have low luminosity so the soft radiation to be dominated by reprocessing and not by the disk thermal emission.  A  luminous disk produces plentiful photons by itself which will linearize the system as they cause steady electron ICS losses that  effectively balance the electron energization. This agrees qualitatively with the observations because QPOs do not appear when  BHXRBs are in the High Soft State -- see, e.g., \cite{remillard2006}.

Our approach aimed to examine the feasibility of the limit cycle occurence when the full energy dependent electron and photon distributions are employed. As key parameters we considered the electron energization timescale $ t_{en}$,
the mass injection rate $\dot{M}_{inj}$
and the radius of the corona $R_c$ where the heating and cooling of the electrons is supposed to take place. Other, auxiliary, parameters include
the reprocessing factor $\alpha$ and the temperature  of the reprocessed photons $T_{eff}$. 
Using a wide range of generally accepted values for the input parameters, with the exception of $ t_{en}$ which we treated phenomenologically, we found that 
the vast majority of the cases shows some type of oscillatory behavior; 
only very fast $t_{en}$ (of the order of $t_{cr}$) and/or very high values of $\dot{M}_{inj}$ do not favor oscillatory solutions. 
Furthermore, 
we found that the frequency of oscillations $\nu_{QPO}\propto R_C^{-1}$ and $\nu_{QPO}\propto {(t_{en}/t_{cr})^{-1/2}}$, see Sect.~\ref{sec:5.3} and \ref{sec:5.1}, respectively, as well as the analytical estimates presented in Sect.~\ref{sec:6},  
while it has a weak dependence on the other input parameters. 

As emphasized earlier, 
in contrast to \citetalias{MPK22}, the present approach can give spectral information about the photon distribution. Taking the average of the output quantities  during a few cycles we found that the X-ray spectra showed either hard indices coupled with high energy spectral peaks or softer indices coupled with lower peaks\footnote{When we refer to spectral indices we mean the ones that are produced simply by ICS. These have not be modified to account for reflection or the Fe-line.}, the corona optical depth in cold electrons was consistently small ($\tau_T<0.1$), while the spectral peaks varied between a few tens to a few hundreds of keV which correspond to electrons reaching maximum Lorentz factors of $\simeq 5-50$. 
Generally, faster acceleration showed spectra with higher peaks and higher bolometric luminosities, as intuitively expected  -- see Tab.~\ref{tab:3} and Fig.~\ref{figtenMW}.  

Another characteristic of the present model is that it can produce both soft (negative) and hard (positive) lags. Appendix~\ref{app2} gives a detailed explanation for this, but, generally speaking, one could associate soft lags to electron cooling, while hard lags come from the contributions of the reprocessed component on the tail of the photons produced from uncooled electrons. The trend is that fast acceleration and low $\dot{M}_{inj}$ tend to produce soft lags, while the situation reverses for slow acceleration and high $\dot{M}_{inj}$ -- see Fig.~\ref{fightDTsh}. This gives the prediction that time lags measured between high energy regimes, where contribution from the disk are not expected, have to be always soft.

The results of the model were compared to the well-studied source GRS 1915+105. We found that the trend observed in the QPO phase lag -- QPO frequency can be satisfactorily reproduced by varying the corona radius $R_c$ while keeping the other input parameters fixed -- see Sect.~\ref{sec:7}. However, as the model is phenomenological, it offers no physical explanation why $R_c$ should change by almost a factor of 10, while $\dot{M}_{inj}$ varies only by a factor of 2. 

Apart from ICS, we have augmented our calculations by using two other relevant physical processes, namely photon-photon pair production and electron-proton bremsstrahlung. These remain marginal for the vast majority of the cases considered. The former can play some role for those cases when the energization pushes the electrons well beyond the GeV range but these require very fast $\tilde t_{en}$ and low $\dot{M}_{inj}$ -- see bottom left corner of Fig.~\ref{fightEpk}.
Generally, since photon compactnesses, as defined in Eq.~\ref{eq:10}, remain well below unity, this process is considered to contribute minimally in the vast majority of the cases. The same can be said for electron-proton bremsstrahlung. 
This process can cause changes in the behavior of the system because it produces radiation that acts as a background photon field, so, practically speaking, it has the same effect as the disk radiation -- see Sect.~\ref{sec:6}. However, its inclusion does not change the essence of our results, at least for the set of parameters examined here.

The present study has neglected the effects of synchrotron radiation. However, as it was shown in Sect.~\ref{sec:6} and App.~\ref{app3}, synchrotron radiations introduces a damping effect and for values of B-field $B\simeq B_{cr}$, see Eq.~\ref{eq:Bmx}, it can impede the oscillatory behavior of the system.
Therefore, all results presented here imply that the B-field of the corona has to be less than $B_{cr}$ which depends on $\tilde t_{en}$ and $R_c$.

Our approach placed emphasis on non-thermal processes because the  compactnesses and the 
associated optical depths of low energy electrons are lower than $0.1$ for most of the parameters studied. For higher compactnesses one should include thermal processes and treat a hybrid thermal/non-thermal plasma \citep{PoutanenCoppi98}. 
However, the non-linearity of the system does not depend so much on the nature of the physical mechanisms used but it is embedded in the equations to be solved. 
Once one treats the system of electrons-photons self-consistenty,  allows the electron cooling to occur on reprocessed photons and includes explicitly an energization method for electrons that is able to replenish the energy lost by the ones which have cooled, then, inevitably, they come to the Lotka-Volterra type of equations which show the limit cycle behavior, as  found both in \citetalias{MPK22} and in the present paper\footnote{For a comparison of the results of the two papers, see Appendix~\ref{app4}}. The only way to damp effectively the oscillations is to allow some physical mechanism which can have a linearizing effect, to operate on comparable timescale to ICS. 

Several papers deal with time-dependent Comptonization to probe the origins of spectral variability and to reproduce the trends of the rms and lags with QPO frequency either in accreting neutron stars \citep{LeeMiller98, KumarMisra14, Karpouzasetal20} or BHXRBs \citep{Karpouzasetal21, Garciaetal21, Bellavita22}. 
Although these works use the same assumptions as ours, they differ in the sense that they assume first the QPO frequency and consequently calculate the quantities of interest. Furthermore, the electron heating rate, when used, is given in each case separately as an external parameter, i.e. they do not use an equation for the electrons which evolves simultaneously with the one for the photons. In our case, we simply write and solve the system of the  electron/photon equations, and there is no guaranty a-priori whether the solution will go to limit cycles or to a steady-state; therefore, there is a conceptual difference of the two approaches in both methods and aims.

The one-zone model employed in the present analysis is the standard framework that is used in modeling the spectra of both AGN/blazars and GRBs, the latter both in the prompt and afterglow phase. It
has the advantage that it can be fully time-dependent and it is able to treat self-consistently the gains and losses of electrons as well as the sources and sinks of photons. The only change that our approach brings to this standard model is to allow for the soft photon feedback. Furthermore, 
we have assumed that electron gains come through a continuous process (see Eq.~\ref{eq:kinetic5}), which we have simply characterized by a timescale $t_{en}$. This allows the electrons to replenish their energy, a feature that is central to our approach. In our case, we have chosen a systematic acceleration of electrons; other approaches that are also possible, but  have not been treated  here, include stochastic acceleration  \citep{Lietal96, StawarzPetrosian08} or an injection term of readily accelerated particles as is usually the approach in blazars \citep{MK97}\footnote{In Appendix~\ref{app5} we have examined the case of Fermi-I type of acceleration timescales without any significant deviations from the basic picture.}. Finally, it should be mentioned that
the present set-up of the problem prohibits the use of $t_{en}\le t_{cr}$. 

The disadvantage of the one-zone model is that it  assumes that both species are uniformly distributed inside the corona, which, in addition, is assumed to be homogeneous and static. Furthermore, it can treat only a spherical geometry and it cannot take into account any details of the disk-corona configuration. Here we have simply assumed that the corona is spherical and it either enshrouds the disk or it is above it, i.e. the lamp-post model. 

The present paper can be used 
to model various black hole systems, both galactic and extragalactic\footnote{In Appendix~\ref{app6} the scaling to the coronae of  Active Galactic Nuclei is briefly presented.} 
 It needs, however, among others, to be implemented with  some physically motivated mechanism for particle energization, it needs to include more radiation mechanisms like synchrotron radiation and, depending on the assumptions, thermal mechanisms.
Other simplifications, like the uniform corona assumed here can be more difficult to address as multi-zone models have not been satisfactorily developed yet. 
However,  the model can accommodate  time-delay effects, i.e. it can allow for a finite time-interval for reprocessing to occur. 
Similarly, the effects of a corona moving outward with a mild relativistic velocity should not, in principle, change the basic concept because  there is still a possibility of a disk-corona feedback \citep{Reig21}.
At any rate, the results of the present analysis show that disk-corona feedbacks are robust and, thus, promising to explain some of the vast BHXRBs phenomenology.

\begin{acknowledgements}
The author would like to thank Dr. S.~Boula for many discussions, substantial support on technical issues and a thorough reading of the final manuscript;  Dr. N.~Kylafis for many discussions and hospitality at the University of Crete where this project was first conceived; Drs. M.~Mendez and M.~Petropoulou for discussions and many helpful comments on the manuscript; 
 Dr. G.~Vasilopoulos for discussions and technical support throughout; and,last but not least, an anonymous referee for their comments that helped improve the present work.
The following Python libraries were used: Numpy \citep{numpy}, Matplotlib \citep{matplotlib}, Scipy \citep{scipy}. For Fig.~\ref{model} we used \cite{gemini}.
\end{acknowledgements}

\bibliographystyle{aa}
\bibliography{references}

\begin{appendix}

\section{A physical interpretation of the limit cycle behavior}\label{app1}

In order to gain a physical understanding of the limit cycle behavior, one needs to follow in detail the spectro-temporal evolution 
of the particle and photon distributions. The exact values of the input parameters are not important for this particular example because, qualitatively, all cases showing oscillatory behavior go through the same phases. However,  we have chosen a case that shows strong limit cycles in order to make the figures clearer.

\begin{figure}
\includegraphics[width = 0.45\textwidth]{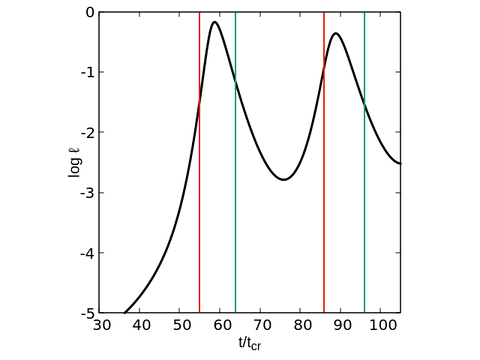}
\caption{A generic light-curve showing the first two peaks of the bolometric ICS compactness. The red and green lines indicate the start of the phases where electron cooling and re-energization prevails respectively -- for details see text.}
\label{figA1}
\end{figure}

Figure~\ref{figA1} shows a close up of the first two limit cycles of a generic case where $\ell_{ds}=10^{-6}$, a value that we used throughout in the main text. The red and green lines separate the time interval according to whether electrons lose or gain energy as this will be evident further down.  Note, however, that these do not coincide with the maxima and the minima of the bolometric photon light curve. We proceed now to analyze this behavior of the light curve by examining in detail the electron and photon distribution functions at each of these time intervals.  \\

\begin{figure*}
\centering
\includegraphics[width = 0.45\textwidth]{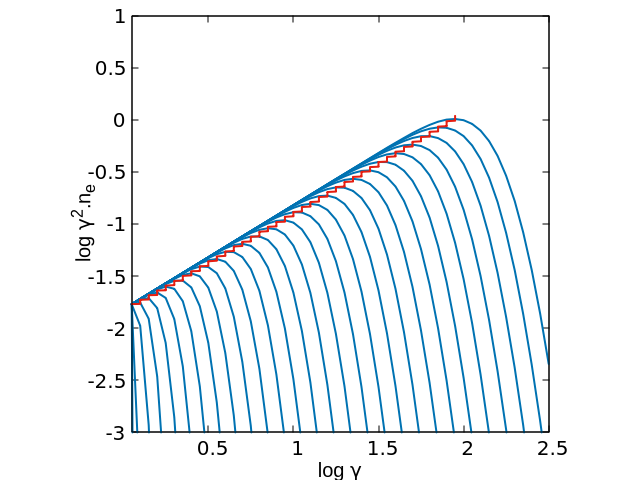}
\includegraphics[width = 0.45\textwidth]{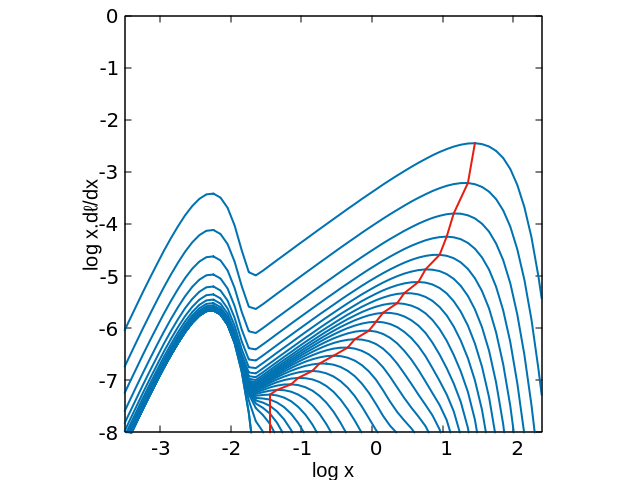}
\caption{Snapshots of a generic case showing the electron distribution $n_e(\gamma)$ multiplied by $\gamma^2$ (left panel) and  the emitted photon spectra in $x.d\ell/dx$ units (right panel) taken every 2$t_{cr}$ during the initial particle energization phase. Here $x={\epsilon}/{m_ec^2}$.
The photon light curve of this case is given in Fig.~\ref{figA1} from $t=0$ up to the first red vertical line.  Red lines in the present figure indicate the peak of the electron instantaneous distribution (left) and of the  ICS photons (right). As time passes both peaks  move from left to right. }

\label{figA2}
\end{figure*}
 \begin{figure*}
\centering
\includegraphics[width = 0.45\textwidth]{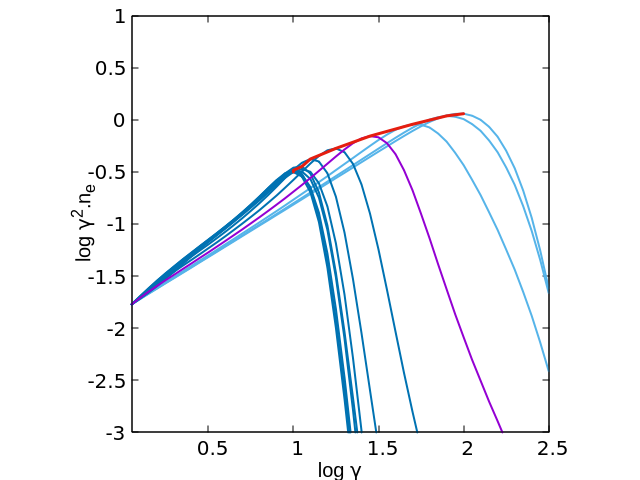}
\includegraphics[width = 0.45\textwidth]{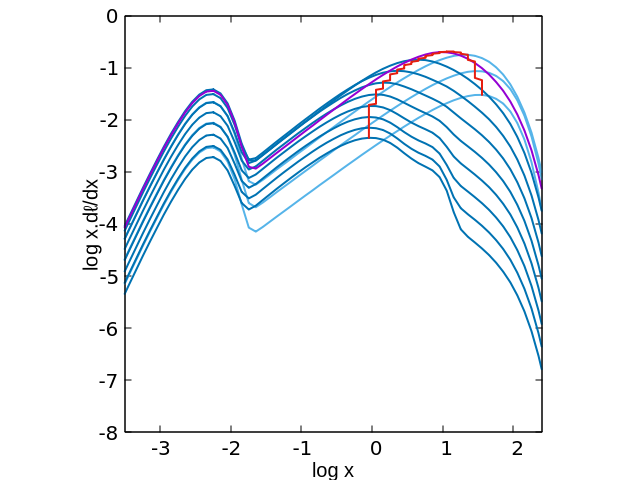}
\caption{Plots of the same quantities as in Fig.~\ref{figA2} for the electron cooling phase, i.e. for the time interval between the red and green lines of Fig.~\ref{figA1}. Snapshots are taken at every $t_{cr}$. The violet lines indicate the distribution functions at the peak of the photon light curve. Light blue lines depict the distribution functions for times before the peak, while the dark blue lines for times after the peak. As time passes, the peaks of both  distributions move from  right to left.} 
\label{figA3}
\end{figure*}

\begin{figure*}
\centering
\includegraphics[width = 0.45\textwidth]{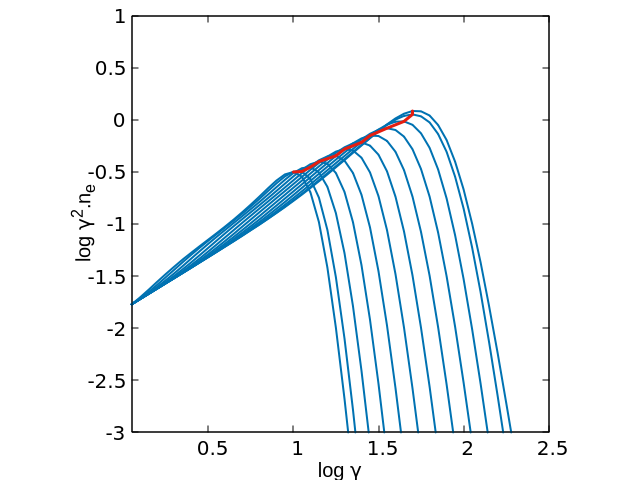}
\includegraphics[width = 0.45\textwidth]{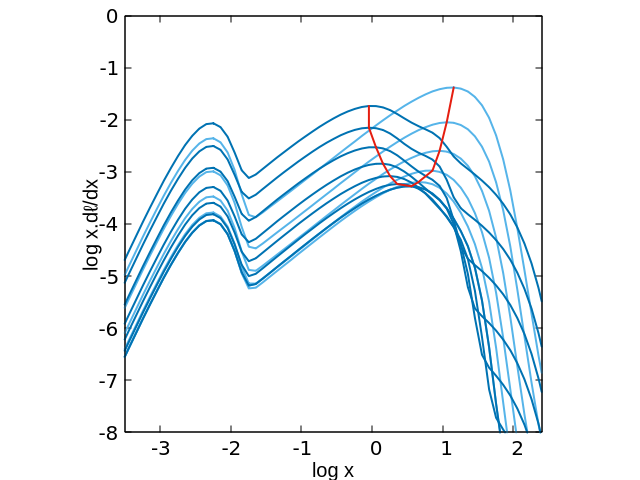}
\caption{Plots of the same quantities as in 
Fig.~\ref{figA2} 
for the electron re-energization phase, i.e. for the time interval defined between the green and the second red line of Fig.~\ref{figA1}. Snapshots are taken at every 2$t_{cr}$.  Light blue lines depict the photon distribution function for times before the photon minimum, while the dark blue lines for times after the minimum. As time passes, the peaks of both  distributions move from left to right.} 
\label{figA4}
\end{figure*}
\textit{1. Initial electron energization phase:}
This phase is physically artificial, because it depends strongly on the initial conditions which have been set equal to zero for both electrons and produced photons. Nevertheless, it has to be included in our analysis because it demonstrates some key features of the model. 

Electrons enter the energization process at low Lorentz factors ($\gamma=\gamma_{inj}$) and,
as time passes, they increase their energy according to Eq.~\ref{eq:2}. This is shown in Fig. \ref{figA2} (left) that draws snapshots of the successive
electron distribution functions, which is drawn  in $\gamma^2n_e$ units\footnote{Here and for the rest of the Appendix, $n_e$ is in dimensionless units obtained by multiplying the original quantity by $\sigma_TR_cm_ec^2$.}, at every 2$t_{cr}$ with the red line connecting their peaks, which occurs at  $\gamma_{pk}(t)$. 
The distributions gradually form a power law which is close to $n_e\propto\gamma^{-1}$ since we have assumed, as in the main text, that $t_{esc}\ll t_{en}$  -- see text below Eq.~\ref{eq:kinetic5}. The corresponding photon spectra, depicted in Fig. ~\ref{figA2} (right) for the same time intervals as the electrons, originally consist only of the disk photons which have a gray-body distribution 
of a very low compactness ($\ell_{ds}=10^{-6}$ by assumption). However, densitydensity $\gamma_{pk}$ increases, they start producing increasingly harder ICS photons, according to the Thomson relation $x_{pk}(t)=\gamma_{pk}(t)^2x_0$, where $x_0$ is the peak of the reprocessed component, the reprocessing of which increases the soft photon density according to Eq.~\ref{equation7}. This positive feedback causes an exponential outgrowth of both hard ICS and reprocessed soft photons. The red line that connects the consecutive peaks of the ICS component moves, therefore, up to higher energies and to higher luminosities, with the former being linear and the latter exponential.
This will last  up to the instant when the soft photons have reached such densities that can cause substantial losses which overcome
the electron energization, i.e. the loss term in the bracket of Eq.~\ref{eq:3} becomes larger than the gain term. From that instant the electrons enter a phase where cooling becomes dominant. This transitional time is shown as a vertical red line in Fig.~\ref{figA1} and it is a universal feature of the present model.\\

\textit{2. Electron cooling phase:}
Figures \ref{figA3} (left) and (right) depict the electron and photon spectral evolution, respectively, during the phase where ICS losses exceed the electron energization gains. Snapshots here are shown at every $t_{cr}$ because the system goes  rapidly through this phase. At  early stages,  the electrons close to the peak suffer severe losses and release their energy producing even more ICS photons, further intensifying their cooling. Thus, electrons cool abruptly (i.e. within a few $t_{cr}$), $\gamma_{pk}$ drops, and, as a result, the rate of ICS and reprocessed photons is reduced. The red line in Fig.~\ref{figA3} is now moving from right to left and as the cooling is gradually reduced electrons become almost stationary at a new, lower $\gamma_{pk}=\gamma_{cool}$. Note that the electron cooling is incomplete, i.e. the soft photon burst cannot force them to cool all way down to their rest mass, so usually $1\ll\gamma_{cool}$. It is also worth mentioning 
that the electron distribution becomes slightly harder due to the cooled high energy electrons that have moved to lower energies causing a pile-up effect. 

The behavior of the photons during this phase is more complex. At its early stages, the photons keep increasing in luminosity (but not in energy) as electron cooling reaches a maximum. The peak of the photon light-curve is typically achieved a few crossing times after the saturation of electrons. The light blue lines in Fig.~\ref{figA3} (right) correspond to the outgrowth of photons at these early stages of the electron cooling phase. These still have a spectral shape similar to the ones during the earlier energization phase. After the photon bolometric peak (depicted here with violet color in both panels) the photon spectra  decrease in energy density and are dominated by (i) photon escape at higher energies 
as the electrons producing them have cooled
and (ii) fresh ICS production at lower energies produced by the cooled electron distribution at the particular time instance -- these spectra are shown in Fig.~\ref{figA3} (right) with dark blue lines.
The red line showing the successive peaks  of the photon distribution (in $x.d\ell/dx$ units) moves in a characteristic counter-clockwise trajectory. The continuous photon escape from their earlier high stages in combination with the low current ICS rate causes the cooling rate to diminish and eventually to become smaller than their energization rate. From then onward, the system moves back to the phase where electrons gain energy. This time instant is shown as the first green  vertical line in Fig. \ref{f_light}. \\

\begin{figure*}
\centering
\includegraphics[width = 0.45\textwidth]{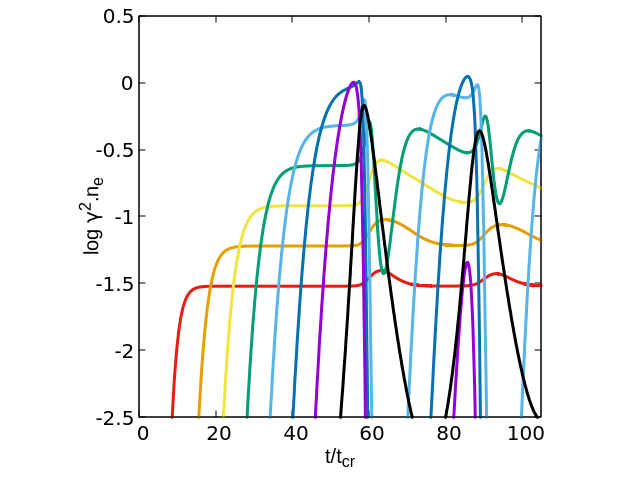}
\includegraphics[width = 0.45\textwidth]{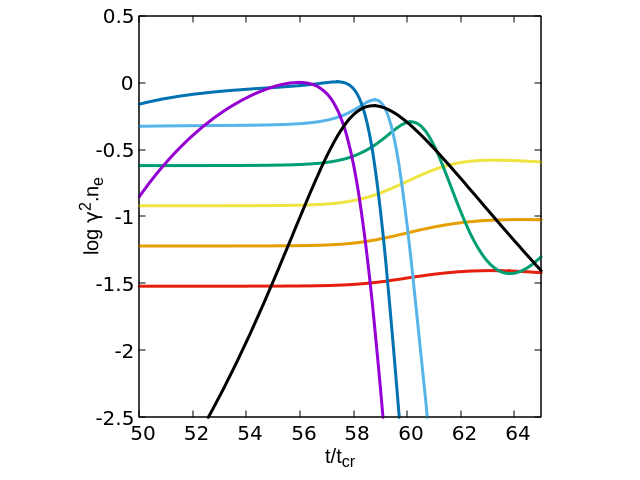}
\caption{Plot of the electron density function $n_e$ multiplied by $\gamma^2$ for various electron energies as a function of normalized time for the run with the parameters that produced the light curve shown in Fig.~\ref{f_MW}. Red line corresponds to $\gamma=10^{0.3}$, orange to $10^{0.6}$, yellow to $10^{0.9}$, green to $10^{1.2}$, light blue to $10^{1.5}$, dark blue to $10^{1.8}$ and violet to $10^{2.1}$. The bolometric photon light curve of Fig.~\ref{f_MW} is shown with black line. The right panel shows a zoom of the first peak for clarity.}
\label{figA5}
\end{figure*}

\textit{3. Electron re-energization phase:}
 Physically, this is very similar to the initial energization phase as the electrons once again move from low to high energies. There are, however, two key differences. (i) Electrons do not start from $\gamma_{inj}$ but from $\gamma_{cool}$, i.e. the value that photons forced them to cool in the previous phase and which depends on the choice of the initial parameters -- see Fig.~\ref{figA5} (left); furthermore, their slope becomes slightly steeper tending again to $p=-1$ as the cooled electrons are re-energized. (b) Early in this phase, higher energy photons continue to drop in luminosity because their spectrum is still dominated by escape -- their spectra are shown in Fig.~\ref{figA5} (right) with dark blue lines. However, gradually a new component of ICS photons produced from the freshly re-energized electrons starts appearing -- shown here with light blue lines. This population dominates at later times and the superposition of these two  (old and fresh) components produces the spectra shown in Fig.~\ref{figA5} (right). The red line depicting the position of the peak photon luminosity moves in a counter-clockwise convex trajectory. As before, this phase will end once ICS photons are built in high enough densities so that the reprocessed photons are able to overcome electron energization and cool them to lower energies. The cycle therefore will repeat all over again.

\begin{figure}
\centering
\includegraphics[width = 0.45\textwidth]{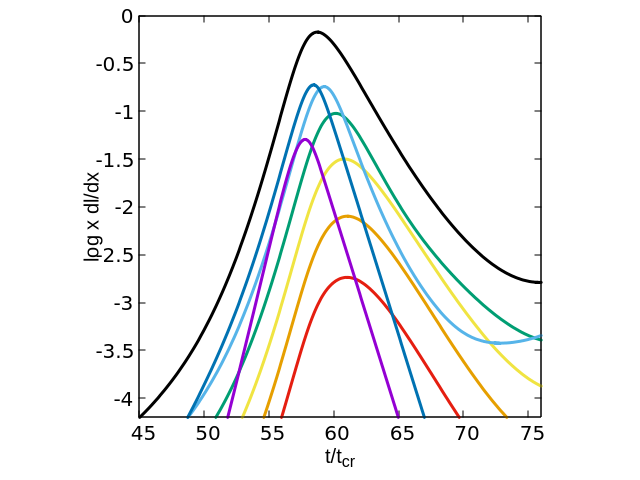}
\caption{Plot of the photon spectrum
 $x.d\ell/dx$ 
 for various photon energies as a function of normalized time around the first peak of the light curve shown in Fig.~\ref{figA1}. The photon energies correspond approximately to the characteristic radiation of the electrons through the ICS relation in the Thomson limit $x_i=\gamma_i^2x_0$. For color codes see Fig.~\ref{figA4}.}
\label{figA6}
\end{figure}

A different way to see the above is  to plot again Figs.~\ref{figA2} - \ref{figA4}, but now  as a function of time, for various characteristic electron and photon energies. Figure~\ref{figA5} plots the quantity $\gamma^2n_e$ vs. normalized time for various electron energies for the same parameters that produced the previous figures. The left panel shows the evolution of the system up to a time as to include the first two photon peaks. As the initial conditions are $n_e(\gamma,0)=n_\gamma(x,0)=0$ one sees  
electrons initially to increase and gradually to obtain a steady state with the lower energies getting there before  the higher ones. Once the soft photons increase and cause electrons to cool, then it is the higher energies that cool faster -- see right panel that shows a zoom of the same electron bins around the first photon peak. Note that it is possible that the higher energies cool before photons reach their peak. Note also that the electrons do not cool completely but only down to a certain energy $\gamma_{cool}$. As already mentioned, this occurs because the ICS cooling timescale depends on energy and photons decrease in density before there is enough time to cool the electrons completely. This can be seen in Fig.~\ref{figA5}: only electron energies with time curves that show dips suffer cooling as their energy is lost to ICS radiation. Therefore, only the cooled electrons play an active role in the photon spectrum and this explains the behavior that one observes in the next figure.  

Figure~\ref{figA6} depicts the ICS photon light curves for seven characteristic energies $x_i$ that correspond to electron energies shown in Fig.~\ref{figA5} through the Thomson relation $x_i=\gamma_i^2x_{0}$.
The light curves are plotted around the first peak of the bolometric ICS luminosity, depicted here with black line. Several characteristic features become readily apparent: Not all photon energies peak at the same time but the lower ones tend to lag from the higher ones. This is to be expected because of the differential cooling of the electrons discussed above. The lag becomes zero at lower photon energies that correspond  to  the uncooled electrons. Therefore, in this example, lags start appearing at the green curve that corresponds to the minimum electron energy that shows some cooling -- see Fig.~\ref{figA5}. Furthermore, it is a small width of energies (less than one order of magnitude) that contribute to the bolometric ICS light curve. Low energies do not contribute because the electrons producing them have not cooled and thus do not radiate  substantially.    

\section{A physical interpretation of time lags}\label{app2}

\begin{figure}
\centering
\includegraphics[width = 0.45\textwidth]{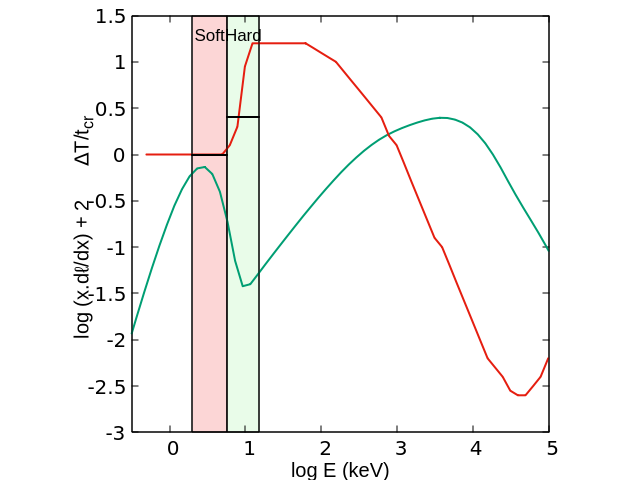}
\caption{Plot of the time lags $\Delta T_i$ of monochromatic photon energy peaks with respect to the time where the black body peak occurs (red line). The average multiwavelength spectrum for the same cycle has also been plotted with green line. The vertical black lines indicate  indicate the "soft" (2-6 keV) and "hard" (6-15 keV) energy bands sometimes used in the literature to measure phase lags. The horizontal black lines indicate the lags in these two bands. In this case the relative soft-hard lag is positive. 
}
\label{figB1}
\end{figure}

The present model shows a rich behavior of soft and hard time lags because the photon spectrum is dominated at high energies by ICS produced from electron cooling, at lower energies by ICS of uncooled electrons, while at even lower energies by the reprocessed soft photons. The two components (ICS and reprocessed soft photons) dominate the spectrum at different energy regimes while there is a rather narrow intermediate energy regime that both contribute partially. 
As an illustrative example we show in 
Fig.~\ref{figB1} the time lags
that have been derived from the first light curve peak for the case examined in Appendix~\ref{app1}.  Here 
the time lags $\Delta T_i$ have been calculated  for the first cycle using the relation
\begin{equation}
\Delta T_i =T_{pk,i}-T_{pk,rpr}
\end{equation}
where $T_{pk,i}$ corresponds to the instant at which the light curve of energy $E_i$ peaks and $T_{pk,rpr}$ corresponds to the 
peak of the reprocessed distribution.
The time resolution is $0.1t_{cr}$ as in the rest of the paper.
For a better understanding of the spectral behavior at the various energy regimes, we  plot in the same figure the average photon spectrum during that cycle. 


\begin{itemize}
\item 
At high energies (in the example here approximately between $10^2$ to $10^4$ keV)
the time lags are soft because the monoenergetic photon lightrcurves are dominated by the eletron cooling -- see Fig. \ref{figA3}. Since the energetic electrons cool faster, one expects that harder photons will appear before softer ones, thus the appearance of a $soft$ lag. Note that the range of photon energies corresponds to $\gamma_{cool}^2x_{0}$ and $\gamma_{pk}^2x_{0}$
where $\gamma_{cool}$ and $\gamma_{pk}$ are the two limiting electron Lorentz factors as shown in Figs~\ref{figA3} and \ref{figA4}
\item 
Below a characteristic photon energy $x_{cool}=\gamma^2_{cool}x_{0}$ the spectrum will be dominated mainly by the tail of the ICS distribution produced by electrons with Lorentz factors $\gamma_{cool}$. These will have a power law of $+2 $ in $xd\ell/dx$ units (see, e.g. \citep{BG70}) and they will oscillate in phase since their radiation is dominated by electrons of Lorentz factors $\gamma=\gamma_{cool}$. Therefore, 
their relative time lags will be zero and the plot will have a $horizontal$ branch. In our example this occurs between   $10$ and $100~\textrm{keV}$. Note also that in the vast majority of the cases examined in the present paper, the peaks of this horizontal branch come after the peak of the reprocessed component. This can be seen in Fig.~\ref{figA6}, where the curves at low energies (red, orange, yellow, green) come after the peak of the bolometric luminosity (black line). This has its significance as we show below. 
\item 
At even lower energies (in the current example for $\epsilon<10~\textrm{keV}$), contributions from the reprocessed  distribution start to become important. Therefore, the spectrum will consist of two components: the tail of the ICS component mentioned above and the Wien-like part of the gray-body spectrum. As photon energies decrease the contribution of the latter becomes gradually more important than that of the former. Thus, since $\Delta T_i>0$ for the light curves of the horizontal part, as mentioned earlier, the time-lag $\Delta T_i$ will show a decreasing trend with decreasing photon energy as it has to reach the value $\Delta T_i=0$ at $x_{0}$. Therefore this part of the spectrum will show a $hard$ lag.
\item 
Finally, it is worth mentioning that at very high energies, there is another hard lag as energization overcomes losses for electrons; however, both electron and photon distributions have very low values there as they are in their high energy cutoffs so one should not expect some observational effect, at least for the parameters used in the present paper.
\end{itemize}

The above analysis covers the basic trends to be expected in a plot of $\Delta T_i$ vs. $\epsilon$. However, there are some variations which depend on the initial parameters. The most common is that for parameters that favor a strong cooling, the cooling energy $\epsilon_{cool}$ moves to the left and, as a result, the horizontal branch will shrink or disappear altogether. In such cases, the soft lag branch gives its place immediately to a hard lag as  the photon energy decreases. 

\begin{figure}
\centering
\includegraphics[width = 0.45\textwidth]{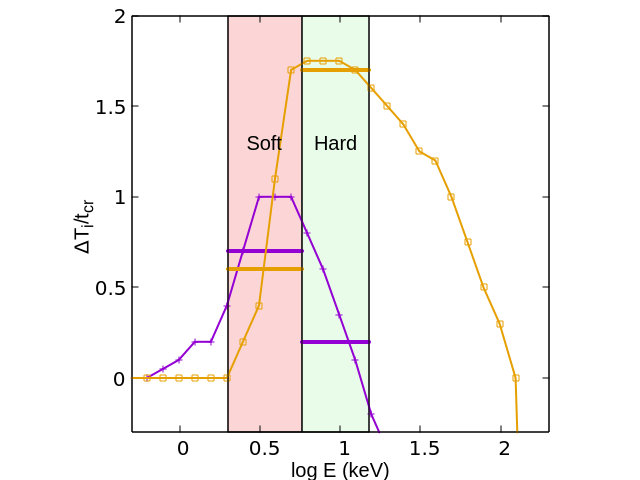}
\caption{Plot of the time lags $\Delta T$ of monochromatic photon energy peaks with respect to the time where the peak of the reflected component occured. The violet line depicts  a negative lag, while the orange one a positive.  The vertical black lines indicate   the "soft" (2-6 keV) and "hard" (6-15 keV) energy bands. The horizontal colored lines indicate the averaged over flux lags in these two bands.  
}
\label{figB2}
\end{figure}

The calculation of $\Delta T_i$ as outlined above allows us to estimate at once the lags between any energy band. For instance, in Fig.~\ref{figB1} we have plotted with vertical black lines the energy bands 2-6 keV (soft) and 6-15 keV (hard) that many authors use to derive lags in QPO observations. By integrating over the corresponding energy bins we can find  
the total energy entering the two bands at each instance and use this to compute time lags (defined as $\Delta T_{sh}$ in the main text) and the corresponding phase lags. Thus, for example, the case examined above shows a hard lag of $\Delta T_{sh}=0.4t_{cr}$. The mean positions of the two lags are depicted with black horizontal lines in Fig.~\ref{figB1}. If, for different parameters, the cooling was increased, then the whole $\Delta T$ curve would have moved to the left and different parts of the curve would have entered the soft and hard bands bringing a change to $\Delta T_{sh}$. This can be seen in Fig.~\ref{figB2} which depicts two such cases. Due to the aforementioned shifting, the orange one shows a positive (hard) lag and the violet one a negative (soft) one.

\begin{figure}
\centering
\includegraphics[width = 0.45\textwidth]{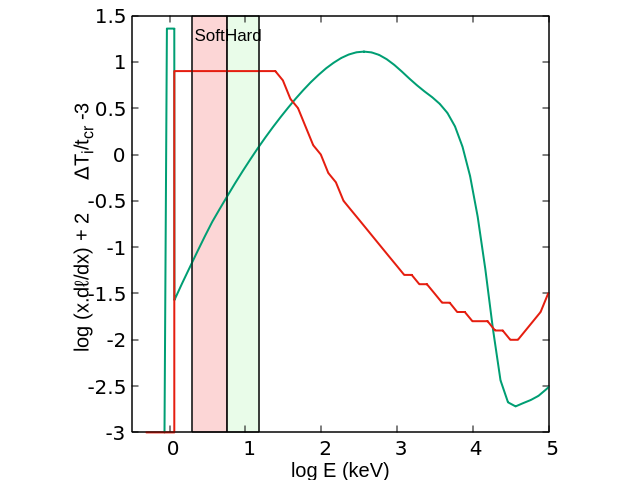}
\caption{Plot of the monoenergetic time lags (red line) and of the multi-wavelength spectrum (green line) in the case where the reprocessed photons are a delta-function.}
\label{figBdelta}
\end{figure}

The fact that hard lags appear in the present model due to the contributions of the Wien part of the gray-body spectrum can be seen in Fig.~\ref{figBdelta} that shows time lags and multi-wavelength spectra as in Fig.~\ref{figB1}, but this time for a delta function reprocessing component. As it can be readily verified, in this case there is no rising part in the $\Delta T$ vs. $E$ plot; instead, 
the so-called "horizontal" branch reaches exactly at the energy of the reprocessing component. 

\begin{figure}
\centering
\includegraphics[width = 0.45\textwidth]{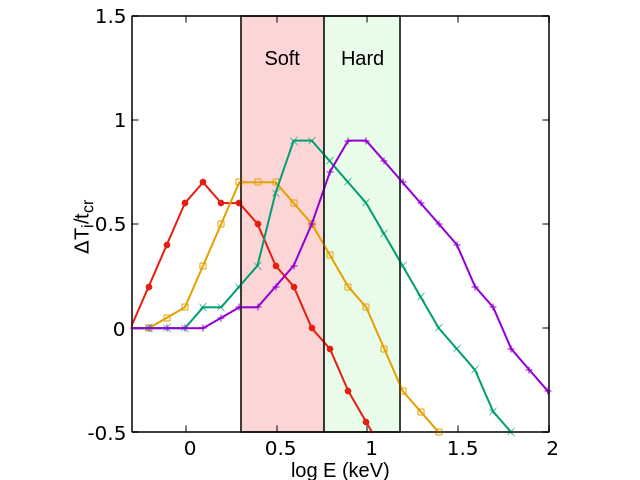}
\caption
{Plot of the time lags $\Delta T$ of monochromatic photon energy peaks with respect to the time where the peak of the reflected component occured. 
The temperatures of this component were assumed to be $10^6~K$ (red), $2\times10^6~K$ (orange), $4\times10^6~K$ (green) and $8\times10^6~K$ (violet) and correspond to the cases shown in Tab.~\ref{tab:7}
of the main text.
The vertical black lines indicate the "soft" (2-6 keV) and "hard" (6-15 keV) energy bands.}
\label{figB3}
\end{figure}

Finally, we should mention that the temperature of the reflected component $T_{eff}$ plays an important role in the sign of the lag. 
Figure~\ref{figB3} compares the time lags when $T_{eff}$ takes different values, from $T_{eff}=1\times 10^6~K$ (red) up to $T_{eff}=8\times 10^6~K$ (violet) which correspond to the runs shown in 
Tab.~\ref{tab:7}
of the main text.
The corresponding curves are shifted with respect to each other, different parts of each enter the soft and hard energy bins, and, as a result, the sign of the lags change from negative to positive as $T_{eff}$
increases. 

The above result can only mean that not only the temperature of the reprocessed component, but its shape as well, might play a role in the sign of the lags. So, for example, one could use a truncated power law instead of a Planck spectrum. However, we will not deal with such cases here, but we note that \cite{ Bellavita22} have reached the same conclusion.

\section{Effects of synchrotron radiation}\label{app3}

As shown in Sect.~\ref{sec:6}, synchrotron radiation can have a damping effect on the limit cycles, provided that the magnetic field exceeds some critical value given by Eq.~\ref{eq:Bmx}. 
This  value, however, is indicative only, as it was derived from solving the simplified system of equations presented in that section.
In the present Appendix we show the effects of synchrotron on the output parameters when the full code is used. 
Tab.~\ref{tab:c} tabulates the usual parameters having as the only variable the corona $B$-field value while keeping the rest parameters  fixed. To determine indicative $B$-values we started from $B_{crit}(\tilde t_{en})$ and we multiplied it by
$~1/4,~1/2$ and $1$. As can be seen from the values of $\tilde T_{10}$ and $f_{rms}$ in the aforementioned  table,  damping sets in practically for $B=B_{crit}/2$ and quenches oscillations for $B=B_{crit}$\footnote{It has to be mentioned here that the exact value of $B$ where damping starts becoming severe depends on the initial parameters. It is safe to assume ,though, that this will be around $B_{cr}.$}. Furthermore, both $L_{bol}$ and $L_{X}$ are reduced with increasing $B$, as synchrotron takes effectively electron energy and channels it to lower photon frequencies, which, however, are efficiently absorbed due to synchrotron self-absorption. Finally, for $B\geq B_{cr}$ synchrotron losses largely impede electron energization, leading to steady-state and low corona luminosity. 

\begin{table*}[h]
  \centering
  \caption{Results as a function the corona magnetic field $B$.}
  \begin{tabular}{ c c c c c c c c c c c}
    \hline \hline
    $B(G)$&    $\nu_{QPO}(Hz)$ & $\tilde T_{10}$ & $f_{rms} (\%)$ & $\Delta T_{sh}/t_{cr}$ & $\Delta\phi$ & $L_{bol}(erg/s)$ & $L_X(erg/s)$ & $E_{pk}(keV)$ & $\Gamma$ & $\eta$  \\ \hline
    
   $0.8\times10^4$ & 1.94 & 1500& 23.8 & -0.04 &-0.02  & $1.7\times10^{38} $&$ 8.4\times 10^{37}$ &280 & 1.3 & 0.009\\ 
    
   $ 1.5\times10^4$ & $1.89$ & 1000& 16.2 & -0.03 & -0.01& $1.4\times 10^{38}$ & $7.9\times 10^{37}$ & 280 & 1.3 & 0.09\\ 

   $3.1\times 10^4$& $1.61$ & 50 & 5.2 & -0.02 & -0.01 & $8.5\times 10 ^{37} $& $5.0\times 10^{37} $& 280 & 1.3 & 0.006 \\

   $ 6.2\times10^4$ & N/A &N/A & N/A & N/A & N/A & $3.0\times10^{35}$ & $6.4\times 10^{34} $& 20 & 2.1 &$2\times10^{-5}$  \\ 

\hline
   
  \end{tabular}
 \tablefoot{Table showing the various output parameters of the problem as a function the corona B-field strength $B$.
 For the definition of these parameters see the caption of Tab.~\ref{tab:3} and text. 
   Other initial parameters are $R_c=10^9~\textrm{cm},~\tilde t_{en}=2.5,~\dot M_{inj}=16\times10^{-8}M_\odot/\textrm{yr},~\alpha=0.1$ and $T_{eff}=4\times 10^6~\textrm{K}$. In the present case $B_{cr}=6.2\times 10^4G$.  }
\label{tab:c}
 \end{table*}

\section{Relation to previous work}\label{app4}

We proceed now to show the relation of the present work to \citetalias{MPK22}. While both works rely on the same principles and essentially come to the same conclusions, they are based on different assumptions.

\citetalias{MPK22} used for powering the electrons a term which was the luminosity related to the mass accretion rate (see their Eq.~\ref{eq:1}). Here, on the contrary, we have 
defined an injected mass accretion rate $\dot{M}_{inj}$ that is not related directly to the luminosity, but was used instead to define the injection rate of electrons in the energization process -- see Eq.~\ref{eq:qinj}. However, since one of our output parameters is the bolometric luminosity $L_{bol}$, we can calculate the mass accretion rate of \citetalias{MPK22} from the relation $\dot m_{MPK}=L_{bol}/L_{Edd}$ and obtain $\dot m_{MPK}$ {\sl{a-posteriori}}. Therefore, this offers a way to compare quantitatively the results of the two papers. Note, however, that for the present setup, the black hole mass is not required; it can be used only indirectly to calculate the expected $L_{Edd}$ and to verify that $R_c>>R_s$, where $R_s$ is the Schwarzschild radius. 

Our numerical results show that both $\nu_{QPO}$ and $L_{bol}$ can be satisfactorily represented as power laws of the energization timescale $\tilde t_{en}$(see Tab.~\ref{tab:3} and Fig.~\ref{fignuten}) and corona radius $R_c$ (see Tab.~\ref{tab:5}). Thus we can write
\begin{equation}
\nu_{QPO}=
C_1\tilde t_{en}^{\alpha_1}R{_c}^{\alpha_2}=
C'_1 t_{en}^{\alpha_1}R{_c}^{\alpha_2-\alpha_1}
\end{equation}
and, similarly, 
\begin{equation}
L_{bol}=
C_2\tilde t_{en}^{\beta_1}R{_c}^{\beta_2}=
C'_2  t_{en}^{\beta_1}R{_c}^{\beta_2-\beta_1}
\end{equation}
where $C_1,~C_1',~C_2,~C'_2$ are proportionality constants; we have also made use of the relation 
$\tilde t_{en}=t_{en}/t_{cr}$.
If we eliminate $ t_{en}$ between the above relations,  we find
\begin{equation}
\nu_{QPO}=C_3L_{bol}^{\alpha_1/\beta_1}R_c^{\alpha_2-\beta_2/\beta_1} 
\end{equation}\label{eq:17}
where $C_3=C'_1C'{_2}^{-\alpha_1/\beta_1}$. Inserting  the values $\alpha_1=-0.49$, $\alpha_2=-1.03$, $\beta_1=-0.85$ and $\beta_2=0.83$ obtained from a $\chi^2-$ fit to the corresponding values tabulated in Tables \ref{tab:3} and \ref{tab:4}, we get $\nu_{QPO}=C_3L_{bol}^{0.57}R_c^{-1.52}$ where $C_3=5.10^{-8}$.
Therefore, Eq.~\ref{eq:17} can be written
\begin{equation}
\nu_{QPO}=5.10^{-8}L_{bol}^{0.58}R_c^{-1.52}~Hz=1.1L_{38}^{0.58}R_9^{-1.52}~Hz,   
\end{equation}
where $L_{38}=L_{bol}/10^{38}~\textrm{erg/sec}$ and $R_9=R_c/10^9~\textrm{cm}$.
This should be compared with the relation (17) of \citetalias{MPK22} where they find $\nu_{QPO}^{MPK}\propto \dot{m}_{MPK}^{0.5}R_c^{-1.5}$. 
Making the correspondence between $L_{bol}$ and $\dot m_{MPK}$, we find that the numerical factors in the two expressions differ by a factor of $\simeq 3$. We note, however, that for this estimation we have used only one curve of Fig.~\ref{fignuten}. Using other curves would change slightly this result.

A difference in the results of the two papers lies on the type of oscillations found. 
In \citetalias{MPK22} the oscillations were always damped, while in the present paper we found that in some cases the oscillations have negligible damping. 
Both papers solve essentially Lotka-Volterra type of equations, albeit with a subtle difference which can be attributed to the implicit assumptions made about electron energy gains.  In \citetalias{MPK22} the energy gains depend on $\dot{m}_{MPK}$, and therefore the electron energy density grows linearly with time (see their Eq.~\ref{eq:2});  in the present approach the energization term used causes the electron energy density to increase exponentially (see Eqs~\ref{eq:2} and \ref{eq:kinetic5} of the present paper as well as Appendix~\ref{app1}). Since the Lotka-Volterra equations initially produce an exponential growth of the preys, the present paper more closely resembles these. However, here we do not solve simply for two equations as in \citetalias{MPK22}, but we use a system of many equations for prey (electrons) and for predators (photons) which complicates the overall picture. These complications can be seen in Appendix~\ref{app1} which shows, for example, that damping starts from photon energies and move towards the lower ones with time.

\section{Energy dependent acceleration and escape}\label{app5}

In the present paper we have adopted an energization and escape timescale that are both independent of energy. However,  in most acceleration scenarios, such as Fermi-I type, one would expect these timescales to increase with electron energy. In this Appendix, therefore, we assume that both energization and escape timescales depend on energy and repeat the procedure outlined in the main body of the paper. 
The formal solution of the electron equation can be found in \cite{KRM98} where it was shown that the electron distribution function $n_e\propto \gamma^{-1-{t_{en,0}/t_{esc,0}}}$ where $t_{en}(\gamma)=t_{en,0}\gamma$ and $t_{esc}(\gamma)=t_{esc,0}\gamma$, therefore the electron distribution function is by one flatter than the energy-independent case studied previously. Table~\ref{tab:D} shows results using the same parameters of Tab.~\ref{tab:3}. All output parameters are comparable, i.e. both $\nu_{QPO}$s and luminosities differ by less than a factor of 2-3, therefore a more realistic choice of the energization timescale does not change the overall results significantly. We note that the luminosities increase in the present case because the electron distribution function becomes flatter, therefore there is even more energy in the upper cutoff.

\begin{table*}[h]
  \centering
  \caption{Results as a function of $\tilde t_{en,0}$.}
  \begin{tabular}{c c c c c c c c c c c}
    \hline\hline
    $\tilde t_{en,0}$ &
    $\nu_{QPO}(Hz)$ & $\tilde T_{10}$ & $f_{rms} (\%)$ & $\Delta T_{sh}/t_{cr}$ & $\Delta\phi$ & $L_{bol}(erg/s)$ & $L_X(erg/s)$ & $E_{pk}(keV)$ & $\Gamma$ & $\eta$  \\ \hline
    1.25 & 3.53 & 30 & 5.7 & -0.13 & -0.10 & $7.1\times 10^{38} $& $4.5\times 10^{38}$ & 360 & 1.6 & 0.046 \\ 
    
    2.5 & 3.17 &500 & 30.0 & -0.1 & -0.06 &$ 4.3\times 10^{38}$ & $2.6\times 10^{37}$ & 360 & 1.6 & 0.024 \\ 
    
    5 & 2.08 & 1500 & 35.0 & -0.1 & -0.04 &$ 2.4\times 10^{38} $& $1.5\times 10^{38} $& 360 & 1.6 & 0.016 \\ 
    
    10 & 1.38 & 10000 & 34.8& -0.17 & -0.05 & $1.1\times 10^{38}$ & $7.6\times 10^{37}$& 230 & 1.7 & 0.007 \\ 
    
    20 & 1.22 & 400 & 21.4 & -0.23 & -0.06 & $3.7\times 10^{37} $& $3.3\times 10^{37}$& 120 & 1.7 & 0.002 \\ 
       
    40 & 1.55 & 30 & 6.7 & -0.4 & -0.13 & $1.7\times 10^{37} $& $ 1.7\times 10^{37}$ & 70 & 1.8 & 0.001 \\ \hline
    `
  \end{tabular}
  \tablefoot{Table showing the values of the various calculated parameters of the problem -- for definitions see captions of Tables 1 and 2)  in the case where $~\tilde t_{en}=\tilde t_{en,0}\gamma$
   as a function of $\tilde t_{en,0}$.
   Other initial parameters are $R_c=10^9~\textrm{cm},~\dot M_{inj}=16\times10^{-8}M_\odot/\textrm{yr},~t_{esc}/t_{en}=10^3,~\alpha=0.1$ and $T_{eff}=4\times 10^6~\textrm{K}$.}

  \label{tab:D}
\end{table*}

\section{Implications for Active Galactic Nuclei}\label{app6}

The present work has been concentrated around QPOs in BHXRBs. However, its scale-free results can be readily applied to Active Galactic Nuclei (AGN) as well. Following the arguments given in Sect.~\ref{sec:5.3} and \ref{sec:6}, one can scale all results obtained in the present work by the following rules
\begin{itemize}
    \item Scale $R_c$ to a new value $R'_c$ and let $\kappa=R'_c/R_c$.
    \item Use a new value of $\dot M'_{inj}=\kappa \dot M_{inj}$.
    \item Keep all other input parameters ($\tilde t_{en}$, $\tilde t_{esc}$, $\alpha$, $T_{eff}$) unchanged. 
\end{itemize}
Then the output parameters will scale as follows
\begin{itemize}
\item The QPO frequency will scale as $\nu'_{QPO}=\kappa^{-1}\nu_{QPO}$.
\item Both $L_{bol}$ and $L_X$ will scale as $L'=\kappa L$.  
\item All other output parameters ($\tilde T_{10}$, $f_{rms}$, $\Delta T_{sh}/t_{cr}$, $E_{pk}$, $\Gamma$ and $\eta$ will remain unchanged.
\end{itemize}
Note that the above show that the results scale with the size of the corona $R_c$ and not with the black hole mass. Based on the above  and on the results shown in Tab.~\ref{tab:3} we can construct Table~\ref{tab:E} for three $R_c$ values that can characterize AGN corona.

\begin{table*}[h]
  \centering
  \caption{Scaling to AGN coronae} \begin{tabular}{c c c c c c c c c c c c}
    \hline\hline
    $R_c(cm)$ & $M_{inj}M_\odot/\textrm{yr}$ &
    $\nu_{QPO}(Hz)$ & $\tilde T_{10}$ & $f_{rms} (\%)$ & $\Delta T_{sh}/t_{cr}$ & $\Delta\phi$ & $L_{bol}(erg/s)$ & $L_X(erg/s)$ & $E_{pk}(keV)$ & $\Gamma$ & $\eta$  \\ \hline

    $10^{14}$ & 0.016 & 
    $1.95\times 10^{-5}$ &1500 & 23.5 & -0.04 & -0.02 &$ 1.7\times 10^{44}$ & $8.5\times 10^{43} $& 280 & 1.2 & 0.011 \\ 

      $10^{15}$ & 0.16 
    & $1.95\times 10^{-6}$ &1500 & 23.5 & -0.04 & -0.02 &$ 1.7\times 10^{45}$ & $8.5\times 10^{44}$ & 280 & 1.2 & 0.011 \\ 
  
      $10^{15}$ & 1.6 
    & $1.95\times 10^{-7}$ &1500 & 23.5 & -0.04 & -0.02 &$ 1.7\times 10^{46}$ & $8.5\times 10^{45}$ & 280 & 1.2 & 0.011 \\ 
  
    \hline

  \end{tabular}
  \tablefoot{Table showing the values of the various calculated parameters of the problem as applied to AGN coronae. Here both $R_c$ and $\dot M_{inj}$ are scaled simultaneously.
   The other initial parameters are $\tilde t_{en}=2.5,~t_{esc}/t_{en}=10^3,~\alpha=0.1$ and $T_{eff}=4\times 10^6~\textrm{K}$.
  
  }
  \label{tab:E}
\end{table*}

\end{appendix}
\end{document}